\documentclass[aps,prb,preprint,groupedaddress]{revtex4}
\usepackage{graphicx} \usepackage{amsmath}

\begin{document}


\title{Optomechanical coupling between a multilayer graphene mechanical resonator
  and a superconducting microwave cavity}



\author{V.~Singh}
\email{v.singh@tudelft.nl}
\author{S.~J.~Bosman}
\author{B.~H.~Schneider}
\author{Y.~M.~Blanter}
\author{A.~Castellanos-Gomez} 
\author{G.~A.~Steele} 
\email{g.a.steele@tudelft.nl}

\affiliation{Kavli Institute of NanoScience, Delft University of Technology, PO Box 5046, 2600 GA, Delft, The Netherlands.}


\date{\today}


\begin{abstract}
\textbf{ The combination of low mass density, high frequency,
  and high quality-factor of mechanical resonators made of two-dimensional crystals such
  as graphene  \cite{bunch_electromechanical_2007,chen_performance_2009,selfcite1,barton_high_2011,eichler_nonlinear_2011,song_stamp_2012,barton_photothermal_2012,chen_graphene_2013}
  make them attractive for applications in force sensing/mass
  sensing, and exploring the quantum regime of mechanical
  motion. Microwave optomechanics with superconducting cavities \cite{regal_measuring_2008,teufel_nanomechanical_2009,rocheleau_preparation_2010,teufel_circuit_2011,massel_microwave_2011,hocke_electromechanically_2012} offers
  exquisite position sensitivity \cite{teufel_nanomechanical_2009} and enables the preparation and detection of
  mechanical systems in the quantum ground state
  \cite{teufel_sideband_2011,palomaki_entangling_2013}. Here, we demonstrate
  coupling between a multilayer graphene resonator with quality factors up to 220,000
  and a high-\textit{Q} superconducting cavity. 
    Using thermo-mechanical noise as calibration, we achieve a displacement sensitivity of 17~fm/$\sqrt{\text{Hz}}$.  
  Optomechanical coupling is demonstrated by
   optomechanically induced reflection (OMIR) and absorption
  (OMIA) of microwave photons
  \cite{agarwal_electromagnetically_2010,weis_optomechanically_2010,safavi-naeini_electromagnetically_2011}. We observe 17~dB of
    mechanical microwave amplification \cite{massel_microwave_2011} and 
  signatures of strong optomechanical backaction.  We extract the
  cooperativity $C$, a characterization of coupling strength,
  quantitatively from the measurement with no free parameters and find
  $C=8$, promising for the quantum regime of graphene motion.}
\end{abstract}

\maketitle




Exfoliated 2-dimensional crystals offer a unique platform for mechanics
due to their low mass, low stress, and high quality factors.
After the initial demonstration of graphene mechanical resonators
\cite{bunch_electromechanical_2007,chen_performance_2009,selfcite1}, there has been a considerable
progress in harnessing the mechanical properties of graphene for mass sensing \cite{chen_performance_2009}, studying nonlinear mechanics \cite{eichler_nonlinear_2011,song_stamp_2012}, and voltage tunable oscillators \cite{barton_photothermal_2012,chen_graphene_2013}. Graphene is also potentially attractive for experiments
with quantum motion: the high quality factor \cite{eichler_nonlinear_2011,barton_high_2011} and low mass of
graphene resonators result in large zero-point fluctuations in a small
bandwidth. The high frequency of graphene resonators also provides low
phonon occupation at cryogenic temperatures. Previously employed
optical \cite{bunch_electromechanical_2007} and radio frequency (RF) techniques \cite{chen_performance_2009,selfcite1,eichler_nonlinear_2011,song_stamp_2012} enabled
detection and characterization of graphene resonators but did not provide a clear path towards the
quantum regime.

A possible route to the ultimate limit in sensitivity is
cavity optomechanics
\cite{aspelmeyer_cavity_2013},
 capable of detecting motion near and below the standard quantum limit
\cite{rocheleau_preparation_2010,teufel_nanomechanical_2009}.
It
has also been used to bring mechanical systems into their
quantum ground state with sideband cooling
\cite{teufel_sideband_2011,chan_laser_2011} and to entangle
microwave photons with the motion of a mechanical resonator
\cite{palomaki_entangling_2013}.
A natural
candidate for implementing cavity optomechanics with graphene resonators is a high-\textit{Q} superconducting
microwave cavity.
Combining graphene with superconducting
cavities in such a way that both retain their excellent
properties, such as their high quality factors, is technologically challenging.

\begin{figure}
\begin{center}
\includegraphics[width=170mm]{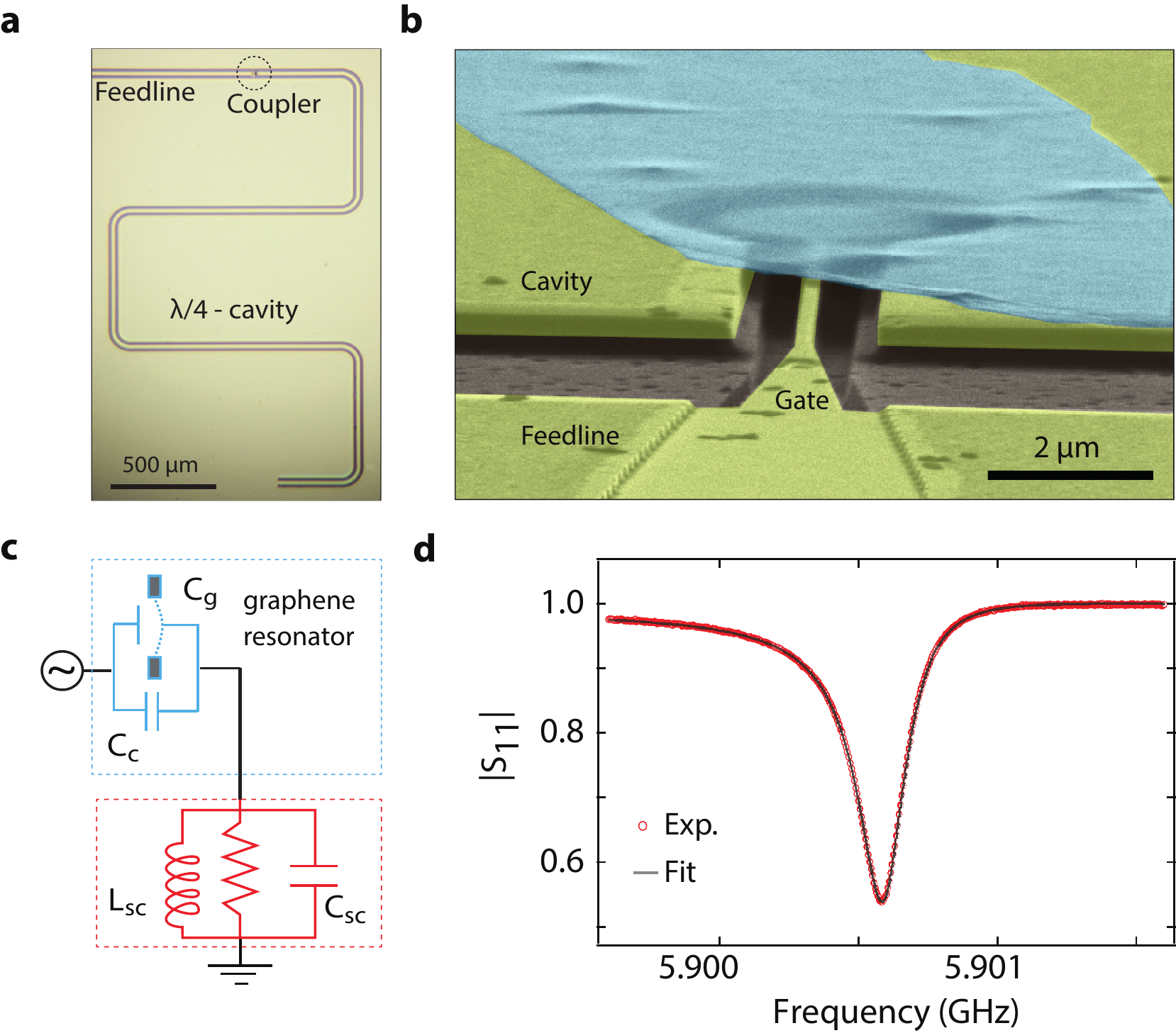}
\end{center}
\caption{\textbf{Characterization of the superconducting cavity coupled to a multilayer graphene mechanical resonator.} (a) Optical image of a superconducting cavity in a quarter wave coplanar waveguide geometry. Microwave photons are coupled in
and out of the cavity via a feedline that is capacitively coupled to
the CPW center conductor by a gate capacitor (coupler). (b) This gate capacitor includes a small gate electrode that extends from the feedline
into the CPW center conductor. A tilted angle scanning electron micrograph (false color) near the coupler showing 4~$\mu$m diameter multilayer (10~nm thick) graphene resonator (cyan) suspended 150~nm above the gate. (c) Schematic lumped element representation of the device with the equivalent lumped parameters as $C_{sc}\approx$ 415~fF and $L_{sc}\approx$ 1.75~nH. (d) Reflection coefficient $|S_{11}|$ of the superconducting cavity measured at $T = 14$~mK (red). Gray curve is a fit to the data, giving the internal quality factor $Q_i\approx107000$ and loaded quality factor $Q_L\approx24250$. \label{fig1}}
\end{figure}

Here, we present a multilayer graphene mechanical resonator coupled to a
superconducting cavity. Using a deterministic all-dry transfer
technique \cite{castellanos-gomez_deterministic_2014} and a novel
microwave coupling design, we are able to combine these two without
sacrificing the exceptional intrinsic properties of either.
Although multilayer graphene has a higher mass than a monolayer, it could be advantageous for coupling to a superconducting cavity due to its lower electrical resistance.
Figures
1(a) and (b) show optical and electron microscope images of the
device. The superconducting cavity is a quarter wavelength
coplanar waveguide (CPW) resonator,
which can be
modeled by an effective lumped capacitance $C_{sc}$ and
inductance $L_{sc}$ with added capacitances from the graphene drum $C_g$ and from the fixed
coupling capacitor $C_c$ to the feedline as shown in Figure~\ref{fig1}(c).

Figure 1(d) shows the measured reflection coefficient ($|S_{11}|$) of
the cavity with low probe power at $T~$=~14~mK. 
A fit of the amplitude and phase of the reflection coefficient yields a 
resonant
frequency $\omega_c\approx2\pi\times5.9006$~GHz, external dissipation
rate of $\kappa_e=2\pi\times$188~kHz and a total dissipation rate of
$\kappa=2\pi\times$242~kHz. The coupling efficiency
$\eta=\kappa_e/\kappa\approx$0.77 ($>1/2$) indicates that the cavity is
overcoupled (See supplementary information (SI) for additional cavity
characterization). 
Based on the designed impedance of the transmission line resonator
($Z_0=51~\Omega$) and geometric
capacitance of graphene resonator $C_{g}\approx$ 578~aF, we estimate
a cavity pull-in parameter of
$G=\frac{d\omega_c}{dx}\approx2\pi\times 26.5$~kHz/nm.

\begin{figure}
\centering
\includegraphics[width=170mm]{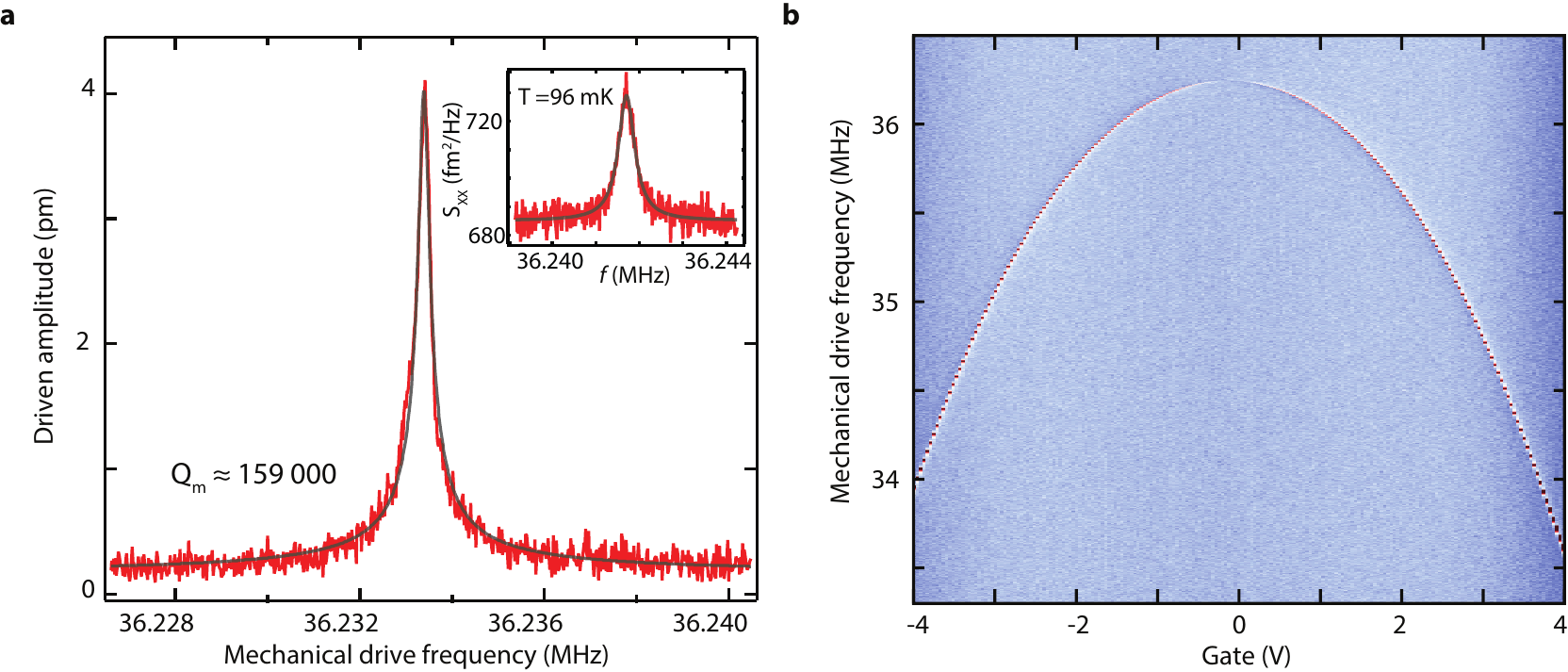}
\caption{\textbf{Sideband resolved detection of the motion.} (a) Driven response of the multilayer graphene resonator at V$_g=$~150~mV (red curve) at $T$~=~14~mK. A Lorentzian fit (gray curve) to the response gives the mechanical quality factor $Q_m\approx159000$, corresponding to mechanical dissipation rate $\gamma_m\approx2\pi\times228$~Hz. Mechanical frequency $\omega_m\approx2\pi\times36.233$~MHz and cavity linewidth $\kappa=2\pi\times242$~kHz implies a sideband resolved limit ($\omega_m\gg\kappa$). The inset shows the displacement spectral density $S_{xx}$ due to thermal noise measured with mechanical mode thermalized at $T$~=~96~mK. The thermal peak corresponds to (27~fm/$\sqrt{\text{Hz}}$)$^2$. (b) Colorscale plot of the homodyne signal with mechanical drive frequency and gate voltage. The sharp change in color represents the resonance frequency of the graphene resonator, indicating its negative tunability with gate voltage. At large gate voltages the response becomes nonlinear due to an increased driving force, which scales linearly with DC gate voltage.  As the electromechanical driving force
($F_{ac}=\frac{dC_g}{dx}V_{ac}V_{g}$) vanishes at zero gate voltage, the
measured signal disappears near $V_g=0$~V.\label{fig2}}
\end{figure}

In Figure 2, we characterize the mechanical properties of the multilayer graphene
resonator using a homodyne measurement scheme \cite{regal_measuring_2008}. Here,
the cavity is used as an interferometer to detect motion while injecting a microwave signal near
$\omega_c$ and exciting the mechanical resonator with an AC voltage applied to the gate.  
Figure~\ref{fig2}(a)
shows the mechanical response of multilayer graphene resonator. A Lorentzian
lineshape fit yields a mechanical resonant frequency
$\omega_m\approx2\pi\times$36.233~MHz and a quality
factor of $Q_m\approx$~159,000 at 14 mK. The mechanical resonance frequency is much larger than the cavity linewidth ($\omega_m/\kappa\sim150$), placing
us in the sideband resolved limit, a prerequisite for ground state
cooling. The cavity can also be used to detect the undriven motion, such as thermomechanical noise of the drum shown in the inset of Figure~\ref{fig2}(a) corresponding to a mechanical mode temperature of 96~mK (see SI for additional details). The thermal motion peak serves as a calibration for the displacement sensitivity.
While driving the cavity at its resonance and utilizing its full dynamic range before the electrical nonlinearity set in (-41 dBm injected power) we estimate a displacement sensitivity for mechanical motion of 17~fm/$\sqrt{\text{Hz}}$.
Using a DC voltage applied to the gate electrode, we can also tune the
frequency of the multilayer graphene resonator shown in Figure~\ref{fig2}(b). The decrease in 
resonance frequency $\omega_m$ for non-zero gate
voltage is due to electrostatic softening of the spring
constant and has been observed before \cite{selfcite1}.

In Figure 3, we demonstrate optomechanical coupling between the
multilayer graphene mechanical resonator and the cavity, mediated by radiation
pressure forces. Figure 3 shows measurement of the cavity similar to
Figure 1(d), but now in presence of a strong drive signal
detuned from the cavity frequency. The frequency is chosen such that
$\omega_d =\omega_c + \omega_m$ $(\omega_c-\omega_m)$, referred as
driving the blue (red) detuned mechanical sideband.  Including this
second detuned drive tone, the cavity response acquires a sharp peak
(Figure 3(a) for the blue sideband) or a sharp dip (Figure 3(b) for the red sideband) centered
at the cavity resonance frequency. The sharp peak in Figure 3(a) when
driving on the blue sideband corresponds to optomechanically induced
reflection (OMIR) of microwave photons, the same phenomenon as
optomechanically induced transparency (OMIT)
\cite{agarwal_electromagnetically_2010,weis_optomechanically_2010,safavi-naeini_electromagnetically_2011},
but now in a reflection geometry. In OMIT, an analogue of
electromagnetically induced transparency (EIT) in atomic physics, a
transparency window opens in the absorption resonance of the cavity
due to the optomechanical coupling to the mechanical
resonator. Driving on the red sideband, Figure 3(b), the cavity shows
a sharp dip in the reflectivity of the cavity, corresponding to a
process of optomechanically induced absorption (OMIA)
\cite{hocke_electromechanically_2012}.

\begin{figure}
\centering
\includegraphics[width=170mm]{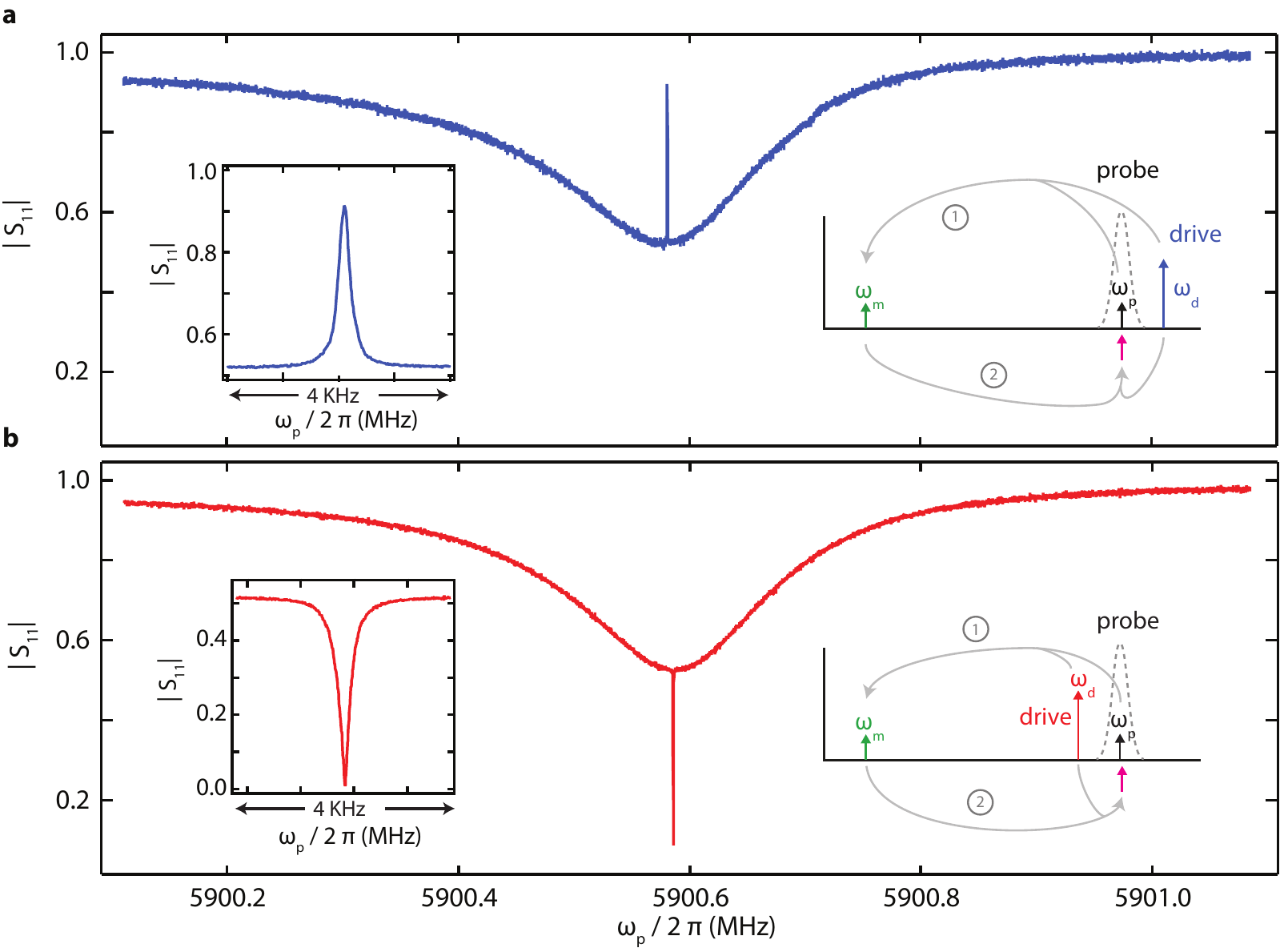}
\caption{\textbf{Optomechanically induced absorption (OMIA) and reflection (OMIR).} Measurement of the cavity reflection $|S_{11}|$ in presence of sideband detuned drive tone at $V_g=0$~V. (a) A blue detuned drive results in a window of optomechanically induced reflection (OMIR) in the cavity response. Inset: Zoom of the OMIR window. (b) A red detuned sideband drive opens an optomechanically induced absorption (OMIA) feature in the cavity response. Inset shows zoom of the OMIA. Schematics in (a) and (b) illustrate the OMIR-OMIA features in terms of the interference of the probe field (black arrow) with the microwave photons that are cyclically down and then up converted by the optomechanical interaction (pink arrow). \label{fig3}}
\end{figure}

Qualitatively, reflection and absorption features in the cavity
response can be understood as follows. Consider the case of blue
detuned driving shown by the schematic in
Figure~\ref{fig3}(a). 
With strong sideband drive signal
 ($\omega_d$, blue arrow) and a weak probe signal ($\omega_p$ near
$\omega_c$, black arrow), the beat between the two microwave signals generates a radiation pressure
force that oscillates at $\Omega\approx\omega_m$ (green arrow), schematically shown by process 1.
This oscillating radiation force at frequency $\Omega$ coherently
drives the graphene resonator. Mechanical motion of the graphene resonator
in turn modulates the drive field resulting in a field appearing at $\omega_d+\Omega$~($=\omega_p$, pink arrow) schematically
shown by process 2. Being a coherent process, the upconverted field
appearing at $\omega_p$ has a definite phase relation with the probe
field and
can either interfere constructively or destructively to produce an
OMIR or OMIA features in the cavity response. In contrast to the first
observation of the OMIT \cite{weis_optomechanically_2010} and other setups \cite{teufel_circuit_2011,zhou_slowing_2013}, here we
observe OMIR on the blue sideband instead of the red sideband, 
a consequence of a overcoupled cavity ($\eta>0.5$) (see SI for detailed discussion).


Using radiation pressure to drive the motion, we can characterize the resonator at
$V_g=0$~V, otherwise inaccessible with
electrostatic driving.
 Using OMIA, we find quality factors as high as
$Q_m\approx$~220,000 at $V_g=0$~V, highest quality factor
detected for a graphene resonator (additional data in the
SI). Inside the OMIA window, a fast change in the phase response
of the microwave signal \cite{safavi-naeini_electromagnetically_2011,zhou_slowing_2013}. This
can be used to implement a microwave photon storage device using
a mechanical resonator (additional data in the SI). For our device, we
estimated a storage time of 10~ms, comparable in magnitude to
that recently demonstrated in conventional mechanical resonators
\cite{zhou_slowing_2013}.
The group delay observed using OMIA here occurs in a very narrow bandwidth, set roughly by the mechanical linewidth, and also with significant optomechanical absorption losses. 

\begin{figure}
\centering
\includegraphics[width=170mm]{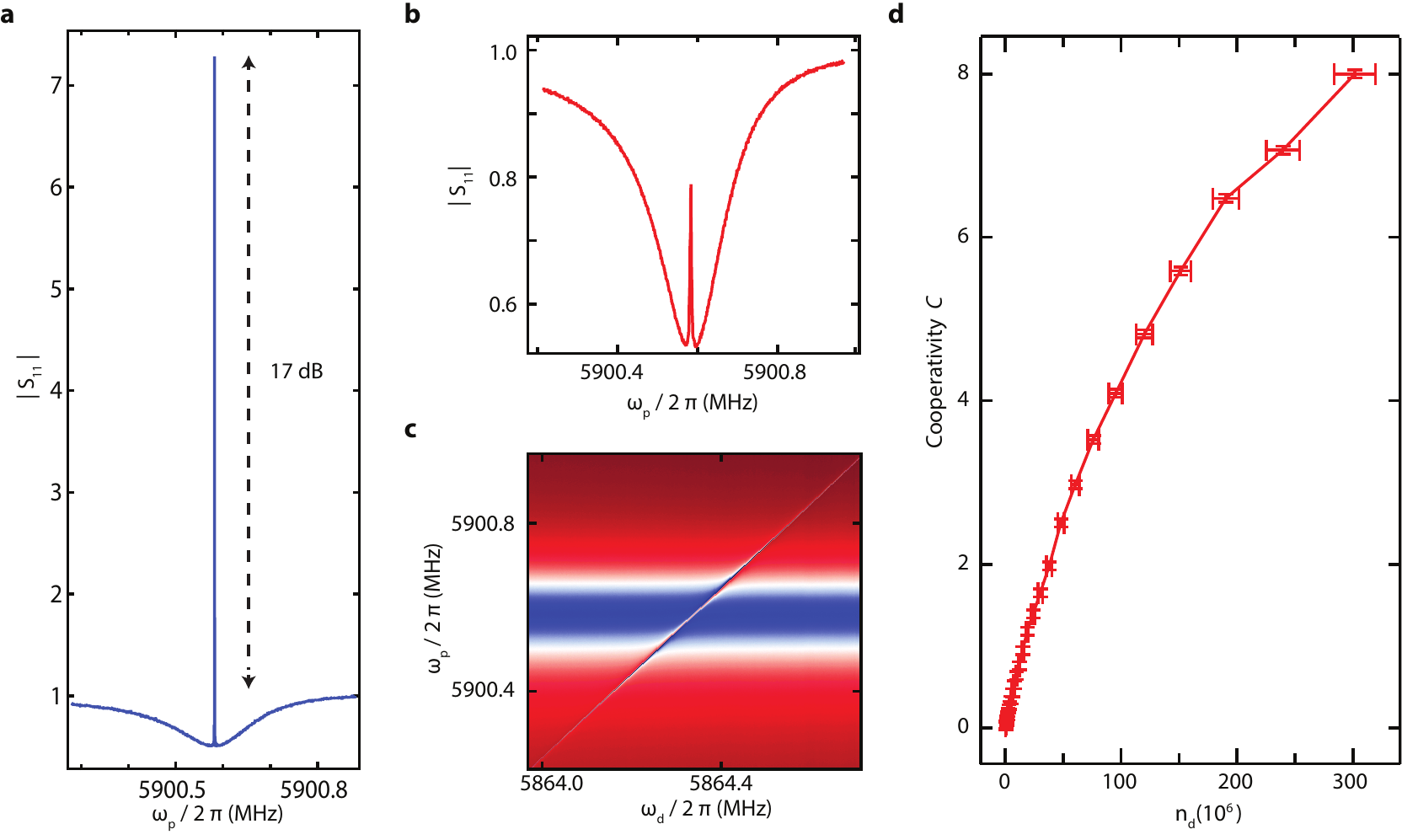}
\caption{\textbf{Large cooperativity with a multilayer graphene mechanical resonator.} (a) Measurement of the cavity reflection $|S_{11}|$ under a strong blue detuned drive ($n_d\approx1.5\times10^7$). At the center of the cavity response, the reflection coefficient exceeds 1, corresponding to mechanical microwave amplification of 17~dB by the graphene resonator. (b) Measurement of the cavity reflection $|S_{11}|$ under a strong red detuned drive ($n_d\approx3\times10^8$). With strong red detuned drive, the OMIA feature becomes OMIR and we observe the onset of the normal mode splitting. Also visible in the colorscale plot in (c) as an avoided crossing while sweeping the drive frequency across the red sideband. The sweep measurement was performed with constant drive power. (d) Plot of cooperativity $C$ vs the number of red detuned photons $n_d$, where we obtained $C=8$. The horizontal error bars result from the statistical uncertainty in determining $g_0,~\kappa~\text{and}~\gamma_m$.\label{fig4}}
\end{figure}

In Figure~\ref{fig4}, we explore the optomechanical interaction with the
multilayer graphene resonator for stronger coupling. The optomechanical
coupling strength ($g$) is tunable by the number of drive photons
inside the cavity ($n_d$) as $g=Gx_{zpf}\sqrt{n_d}=g_0\sqrt{n_d}$,
where $x_{zpf}$ is the amplitude of quantum zero-point fluctuations of the mechanical
resonator and $g_0$ is the single photon coupling strength
\cite{aspelmeyer_cavity_2013}. We calibrate $g_0$ using a frequency modulation scheme \cite{gorodetsky_determination_2010,zhou_slowing_2013} (additional data in SI) and find $g_0=2\pi\times0.83$~Hz.
The combination of the calibration of $g_0$ together with the OMIA data allows an absolute calibration of photon number in the cavity. Figure~\ref{fig4}(a) shows the cavity
response while driving at the blue sideband with
$n_d\approx1.5\times10^7$. Compared to the reflection feature show in
Figure~\ref{fig3}(a), the OMIR peak has increased such that the
reflection coefficient now exceeds unity, indicating more probe
photons are coming out than we are putting in.  This gain of microwave
photons arises from mechanical microwave amplification
\cite{massel_microwave_2011} by the graphene resonator. We
achieve up to $\sim$17~dB gain in a bandwidth of $\sim$ 300 Hz with a number of added photon noise
that can theoretically be as low as the number of thermal phonons ($\bar{n}_{th}$) in
the mechanical resonator\cite{massel_microwave_2011}. Increasing the blue detuned drive further leads to the self oscillation of the mechanical resonator. The origin of microwave amplification can also be
understood in terms of the optomechanical backaction on the cavity:
when the coupling strength becomes sufficiently strong such that
$4g^2/\gamma_m$ overcomes internal dissipation rate of the cavity
$\kappa_i$, the net effective internal damping
rate of the microwave cavity becomes negative.

Figure~\ref{fig4}(b) shows the cavity response for a red detuned drive
with $n_d\approx3\times10^8$.  In contrast to Figure~\ref{fig3}(b), the absorption feature has become a reflection (transparency)
window.  This crossover occurs when the optomechanical coupling becomes
sufficiently strong such that effective internal dissipation rate of
the cavity $\kappa_i^{eff}=\kappa_i+\frac{4g^2}{\gamma_m}$ exceeds the external
dissipation rate $\kappa_e$.
The reflection window in
Figure~\ref{fig4}(b) is considerably larger in width than in
Figure~\ref{fig3}(a). This is an indication of optomechanical
backaction: Similar to the modification of
the internal dissipation rate of the cavity, the mechanical
dissipation rate ($\gamma_m$) is also modified, $\gamma_m^{eff} = \gamma_m\left(1+\frac{4g^2}{\kappa\gamma_m}\right)$, leading to the
broadening of the OMIR feature and onset of the normal mode splitting of the cavity resonance \cite{teufel_circuit_2011,weiss_strong-coupling_2013}.
At the highest photon numbers, we achieve $\frac{2g}{\kappa}\sim~0.12$,
corresponding to 12\% of the strong coupling limit. At these coupling rates, a weak avoided crossing in the cavity response can be seen when a detuned drive is swept across the red sideband, shown in
Figure~\ref{fig4}(c).

A useful feature of OMIA-OMIR data is that height of the peak or
depth of the dip allows one to extract
$\frac{4g^2}{\kappa\gamma_m}\equiv C$, the optomechanical
cooperativity, with no free parameters. The cooperativity $C$ is an
important characterization for the optomechanical experiments. For example, in sideband
resolved limit, the criteria for cooling the mechanical
resonator to its quantum ground state is $C+1 >
\bar{n}_{th}$ , where $\bar{n}_{th}$ is the number of thermally excited 
quanta in the resonator \cite{aspelmeyer_cavity_2013}. Figure~\ref{fig4}(d) shows the
cooperativity of our device extracted from the OMIA data as a function
of the number of red detuned drive photons inside the cavity
($n_d$). Increasing $n_d$, we have been able to achieve
$C=8$.
Although $C$
should scale with $n_d$, in our device the intrinsic mechanical
quality factor begins to decrease at higher cavity power leading to
deviation from linear relation
(see SI for detailed characterization). 
With $C~=~8$, we achieve a maximum multi-photon coupling
strength $g\sim2\pi\times14$~kHz with $n_d\approx3\times10^8$.

We conclude by discussing future prospects of the optomechanical device presented here. 
In principle, assuming an equilibrium with the phonon bath at 14 mK, cooperativity achieved here brings the system close to the quantum coherent regime $C\sim\bar{n}_{th}=\frac{k_BT}{\hbar\omega_m}$. 
In practice, however, a higher margin on cooperativity is desired in order to overcome considerations such as technical noise or mode heating at high powers.
Our estimates suggest that $C/\bar{n}_{th}$ will scale strongly with area indicating that significant gains can be made with larger area drums (see SI for details).
Quantum applications aside, using our graphene optomechanical device, we have demonstrated 17~dB microwave
amplification, potential for microwave photon storage for up to
10~ms, and observed a record-high mechanical
quality factor $Q_m\sim220,000$ for a crystalline multi-layer graphene resonator using radiation pressure force
driving. 
Our cavity as a detector offers a displacement sensitivity of 
17~fm/$\sqrt{\text{Hz}}$ with a bandwidth 3 orders of magnitude larger
than the mechanical dissipation rate ($\kappa/\gamma_m\sim10^3$). 
This
provides an unprecedented new tool for studying phenomena as nonlinear restoring forces, nonlinear damping, and
mode coupling in graphene. 
Finally, the combination of our microwave design and transfer technique for resonator fabrication can be easily extended to other exfoliated 2D crystals, including supeconducting materials like NbSe$_2$, potentially eliminating any resistive losses in the coupled optomechanical system.

\subsection*{Methods}

We have used Mo/Re alloy to fabricate the superconducting
cavity. Large superconducting transition temperature ($T_c\sim8$~K) allows us to
achieve large dynamic range of the cavity. To fabricate the quarter
wavelength superconducting cavity in coplanar waveguide geometry, we
first clean the intrinsic Si substrate with hydrofluoric acid. These
substrates were immediately loaded for sputtering and a layer of
300~nm Mo/Re was deposited. Following standard e-beam lithography
(EBL) and reactive ion etching procedures, we etched out a quarter wavelength
cavity.
The gate is thinned down
in a second etch step so that it is 150 nm below the surface of the
CPW center conductor metal. In the last step of fabrication, a
multi-layer graphene flake (thickness $\sim 10$ nm) is transferred
over the hole, forming a capacitor between the superconducting CPW
resonator and the feedline.

In this scheme, the motion of graphene dispersively
modulates the cavity frequency $\omega_c$ through the change in
$C_g$ : $\omega_c(x)=1/\sqrt{L_{sc}(C_{sc}+C_c+C_g(x))}$~~\cite{dissipative}. The use of the feedline capacitance to couple to
the graphene resonator has some convenient practical advantages. In particular DC
gate voltages (for tuning the mechanical frequency) and low frequency RF
voltages (for electrostatically driving the mechanical resonator) can easily be applied
to the microwave feedline without spoiling the quality factor of the
superconducting cavity.



\subsection*{Acknowledgments}

Authors would like to thank Aashish Clerk for providing initial theoretical input. This work was supported by the Dutch
Organization for Fundamental Research on Matter (FOM). A.C.G. acknowledges financial support through the FP7-Marie Curie project
PIEF-GA-2011-300802 (`STRENGTHNANO')

\subsection*{Author contributions}

V.S., S.B. and B.S. optimized and fabricated the quarter wavelength
superconducting cavity (SC) samples. A.C.G. developed the
deterministic transfer method. V.S. and A.C.G. transfered graphene on
the SC samples. V.S. and S.B. set up the low temperature microwave
measurement setup. V.S. performed the measurement. Y.B. provided
theoretical support. G.A.S. conceived the experiment and supervised
the work. All authors contributed to writing the manuscript and give
critical comments.

\subsection*{Additional information}

Supplementary information is available in the online version of the
paper. Reprints and permissions information is available online at
www.nature.com/reprints. Correspondence and requests for materials
should be addressed to V.S. or G.A.S.

\subsection*{Competing financial interests}
The authors declare no competing financial interests.

\newpage

\end{document}



\title{Supplementary Information: Optomechanical coupling between a multilayer graphene mechanical resonator
  and a superconducting microwave cavity}



\author{V.~Singh}
\email{v.singh@tudelft.nl}
\author{S.~J.~Bosman}
\author{B.~H.~Schneider}
\author{Y.~M.~Blanter}
\author{A.~Castellanos-Gomez} 
\author{G.~A.~Steele} 
\email{g.a.steele@tudelft.nl}



\date{\today}

\maketitle

 
 
%
\makeatletter 
\def\tagform@#1{\maketag@@@{(\ignorespaces#1\unskip\@@italiccorr)}}
\makeatother
 

\makeatletter
\makeatletter \renewcommand{\fnum@figure}
{\textbf{Supplementary~Fig}.~\thefigure}
\makeatother
 
\renewcommand{\bibnumfmt}[1]{\text{[S#1]}}
\renewcommand{\citenumfont}[1]{\text{S#1}}
 


\section{Estimation of pre-tension in the mechanical resonator $T$:}

The resonant frequency of the fundamental mode for a circular resonator in the plate limit, is given by \cite{timoshenko_vibration_2007},
\begin{eqnarray}
f_{plate} = \frac{10.21}{\pi}\sqrt{\frac{E}{3 \rho (1-\nu^2)}}\times\frac{t}{d^2}
\end{eqnarray}
where, $E$ is the Young's modulus of rigidity, $\nu$ is the Poisson's ratio, $\rho$ is the volume mass density, $t$ is the thickness and $d$ is the diameter of the resonator. In the other extreme limit, when pre-tension ($T$) in the flake dominates over the bending rigidity, the resonant frequency is given by \cite{timoshenko_vibration_2007},
\begin{eqnarray}
f_{tension} = \frac{2.4048}{\pi d}\sqrt{\frac{T}{\rho t}}\label{eq:preT}.
\end{eqnarray}
In the cross-over limit, when these two contributions are comparable, the two limits can be joined by,
\begin{eqnarray}
f_m = \sqrt{f_{plate}^2 + f_{tension}^2} \label{eq:joinf}
\end{eqnarray} 
Using $\rho = 2200~$kg/m$^3$, $E=1.02$~TPa, $\nu=$~0.165, $t=10$~nm and $d=4~\mu$m, we get $f_{plate}=$~25.6~MHz, which is smaller than the measured resonant frequency of the device ($\sim$~36.213~MHz), suggesting the additional contribution from the tension in the resonator. Using equations~\ref{eq:preT} and \ref{eq:joinf}, we calculate a pre-tension of $T = 0.39$~N/m (pre-stress = 39 MPa), which is comparable to that observed in other suspended few layer graphene flakes with similar geometry \cite{poot_nanomechanical_2008}.
It should be noted that here in our device the ratio of mechanical compliances (restoring force) due to pre-tension and the bending rigidity $\frac{k_{stress}}{k_{bending}}$ is of the order of unity. This is very small compared to the ratio observed in high-$Q$ SiN resonators,
where ratios on the order of $10^8$ are needed to achieve high-$Q$ \cite{unterreithmeier_damping_2010}

\subsection{Capacitance of the drum resonator $C_g$ and cavity pull-in parameter $G$ :}

Using finite element simulation for capacitance, we find the graphene drum capacitance to be $C_{g}\sim$~578 aF (for a 4~$\mu$m diameter graphene drum suspended over 3~$\mu$m diameter gate with 150~nm air gap). Using the cavity lumped parameters as $C_{sc}\sim$~415~fF and $L_{sc}\sim$~1.75~nH and assuming a parallel plate capacitor model, we get the cavity pull-in parameter $G=\frac{d\omega_c}{dx}=2\pi\times26.5$~KHz/nm.

\subsection{Quantum zero point fluctuations $x_{zpf}$ and single photon coupling strength $g_0$ :}

To estimate the single photon coupling strength, we have used the total mass of the resonator ($m=\pi (d/2)^2 t \rho =0.276$~pg) giving quantum zero point fluctuations $x_{zpf} = \sqrt{\frac{\hbar}{2m\omega_m}}\sim$~29~fm. This gives us an estimate on single photon coupling strength rate based on the geometric modeling  $g_0=Gx_{zpf}\sim2\pi\times0.76$~Hz. We have performed an effective mass independent calibration of $g_0$ using a frequency modulation technique (discussion in the coming section). This calibration results in $g_0 = 2\pi\times0.83$~Hz.
Note that the calibration procedure allows one to extract $g_0$ without making any assumptions about the definition of mode amplitude or of effective mass. Such an assumption is, however, necessary when calculating the displacement sensitivity from the calibrated $g_0$. For this, we will chose the convention of Poot \textit{et. al.}\cite{poot_mechanical_2012} taking the effective mass as the total mass. Such a convention normalizes the peak amplitude based on the mode-shape.

\section{Comparison between dispersive and dissipative coupling strength :}

The dimensionless dispersive $\tilde{A}$ and dissipative $\tilde{B}$ coupling strengths are given by \cite{elste_quantum_2009},
\begin{eqnarray}
\tilde{A} = \frac{1}{\kappa}\frac{d\omega_c}{dx}x_{zpf} && \text{and ~~} \tilde{B} = \frac{1}{\kappa}\frac{d\kappa}{dx}x_{zpf}
\end{eqnarray}
where, $\omega_c$ is the cavity resonant frequency, $\kappa$ is total dissipation rate, and $x_{zpf}$ is amplitude of the quantum zero point fluctuations of the mechanical resonator. Utilizing the lumped element model as described in the main text, the ratio between $\tilde{A}$ and $\tilde{B}$ can be written as,
\begin{eqnarray}
\frac{\tilde{B}}{\tilde{A}}\simeq 2 \frac{\kappa}{\omega_c}\frac{C_{sc}}{C_{e}}
\end{eqnarray}
where, $C_e = C_g+C_c$ is total coupling capacitance between microwave feedline and superconducting cavity.
Using the parameters of the device described in the main text, we obtain $\left(\frac{\tilde{B}}{\tilde{A}}\right)^2\approx3\times10^{-4}$. Therefore in comparison with the dispersive coupling, the dissipative coupling can be ignored.

\section{Cavity characterization :}

The reflection from a single port cavity can be described by \cite{aspelmeyer_cavity_2013},
\begin{eqnarray}
S_{11}=\frac{\frac{\kappa_i-\kappa_e}{2}-i(\omega-\omega_c)}{\frac{\kappa_i+\kappa_e}{2}-i(\omega-\omega_c)}
\end{eqnarray}
where $\kappa_i$ and $\kappa_e$ are the internal and external dissipation rate of the cavity. It is interesting to note that at the resonant frequency the reflection coefficient $S_{11}^0 = |S_{11}(\omega=\omega_c)|=|1-2\frac{\kappa_e}{\kappa}|$ is determined by the coupling efficiency $\eta\equiv\frac{\kappa_e}{\kappa}$. 

\begin{figure}
\begin{center}
\includegraphics[width=140mm]{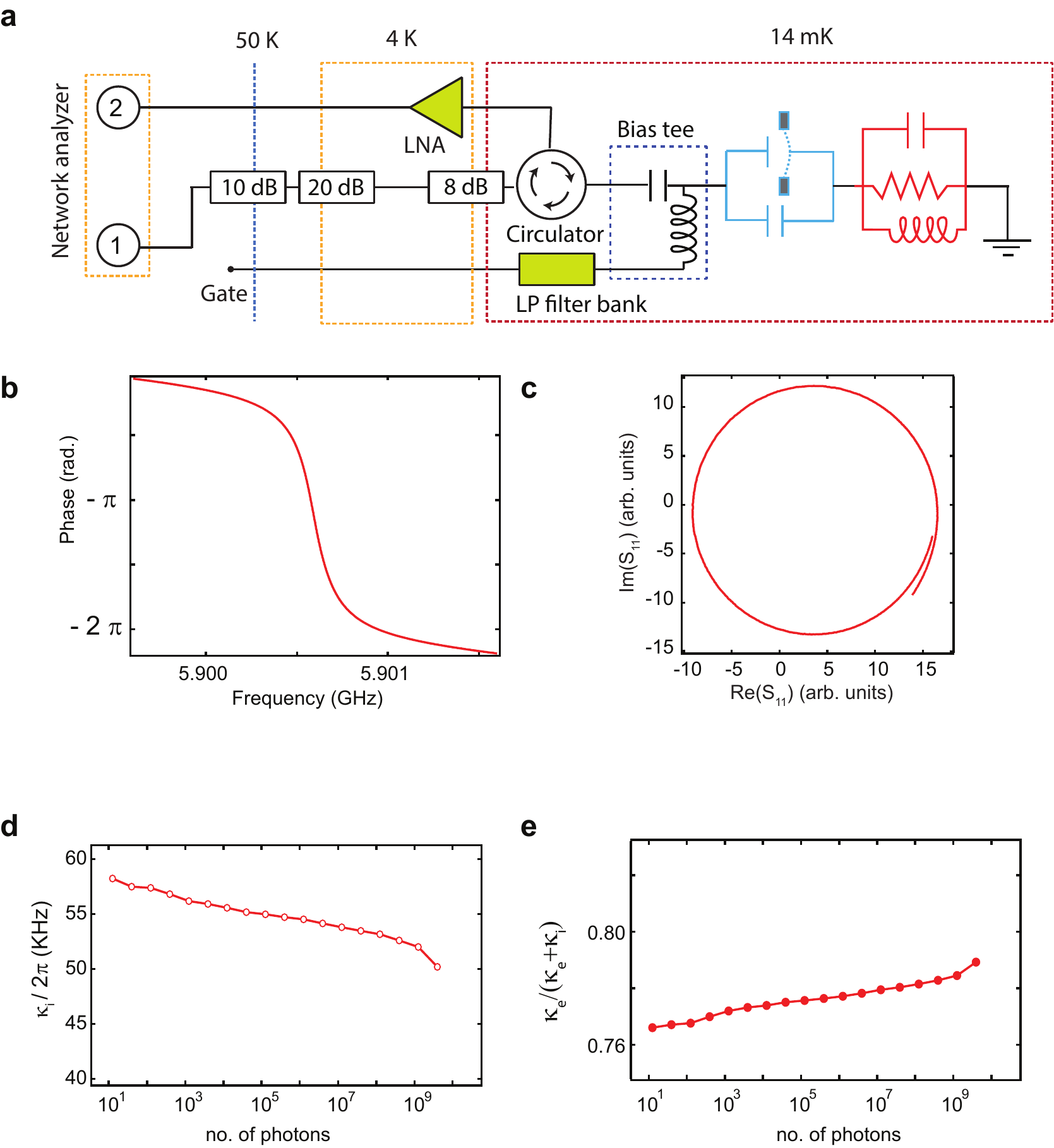}
\end{center}
\caption{a) Schematic of the complete measurement setup. The attenuation between 4K and 14 mK is configured as 6 dB at an 800 mK stage, 1 dB at a 50 mK stage, and 1 dB at the 14 mK stage. We note that this is an insufficient attenuation to fully thermalize the microwave field, resulting in a microwave photon temperature of about 800 mK. b) Arg($S_{11}$) measured across the resonance indicating a swing larger than $\pi$ radian, which is a characteristic of an overcoupled single port reflection cavity. c) X and Y quadrature plotted against each other across the resonance. d) Plot of the internal decay rate $\kappa_i$ of the cavity with the number of probe photons inside the cavity. e) Plot of coupling fraction $\frac{\kappa_e}{\kappa_e+\kappa_i}$ with number of probe photons inside the cavity. \label{SIfig1}}
\end{figure}

Supplementary Fig.~\ref{SIfig1}(a) shows the schematic of the complete measurement setup used for cavity characterization. From a vector network analyzer, a microwave signal is sent to the device through highly attenuated cables. A cold circulator is used to measure the reflection from the cavity. Reflected signal is then amplified by a low noise amplifier (LNA) with a noise temperature specified to be 2.65~K (Low Noise Factory LNF4-8). To apply DC gate voltage on the microwave feedline, we use a separate DC line (highly filtered with low pass RC and copper power filter) and a bias tee to add it to the microwave feedline. 

Supplementary Fig.~\ref{SIfig1}(b) shows the plot of the phase of $S_{11}$ measured across the resonance. A swing of more than $\pi$ radian shows the overcoupled nature of the cavity. Overcoupled behavior of a single port reflection cavity can also be seen by plotting X and Y quadratures of $S_{11}$ against each other (resonance circle). An overcoupled behavior reflects as a complete circle enclosing the origin as shown in Supplementary Fig.~\ref{SIfig1}(c).

Supplementary Fig.~\ref{SIfig1}(d) shows the plot of the internal decay rate of the cavity with varied number of probe photons. As it can be seen from the figure, with large number of the probe photons, internal decay rate reduces. Supplementary Fig.~\ref{SIfig1}(e) shows the plot of coupling efficiency $\eta=\frac{\kappa_e}{\kappa_e+\kappa_i}$ with number of photons. As the number of probe photons becomes large, $\kappa_i$ reduces and hence enhances the coupling efficiency.

As the response of the cavity has a Fano lineshape, (possibility arising from the finite isolation in the cold circulator), we fit the response by considering  a complex  $\kappa_e=|\kappa_e|e^{i\phi}$. Alternatively, a finite isolation from the circulator, the measured reflection form the cavity can be written as,
\begin{align}
S_{11}(\omega) = \alpha e^{i\phi} + (1-\alpha)(1-\frac{\kappa_e}{-i\Delta+\kappa/2}) 
\end{align}
where, $\alpha$ is isolation from port 1 to port 3 of the circulator and other symbols have their usual meanings. This approach yields less than 1 percent difference on the cavity dissipation rates estimate calculated by taking a complex $\kappa_e$.

\subsection*{Tuning the cavity frequency with DC gate voltage:}

\begin{figure}
\begin{center}
\includegraphics[width=100mm]{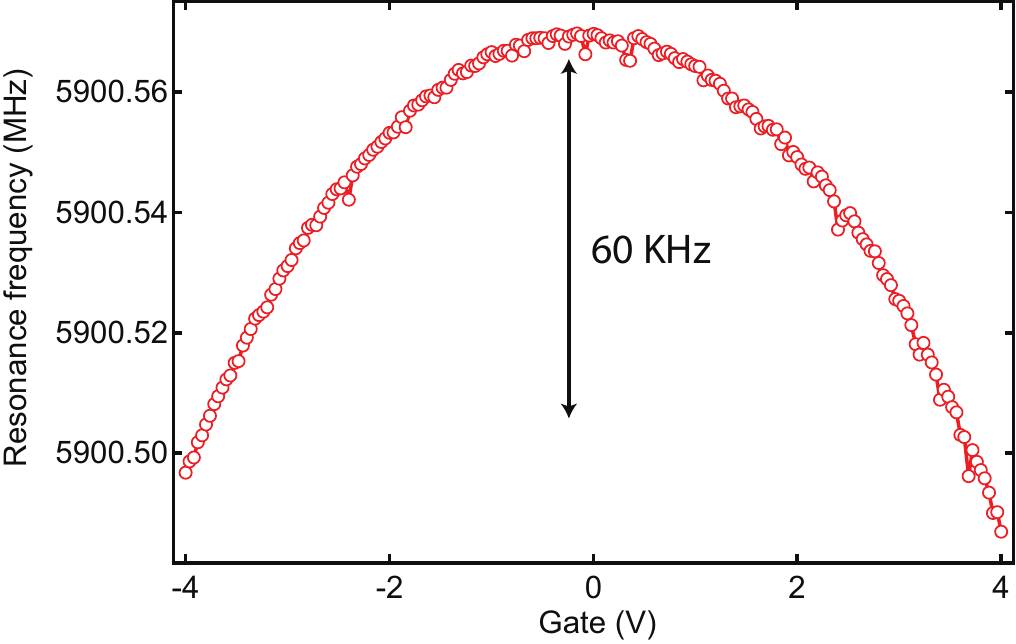}
\end{center}
\caption{ Change in the cavity resonance frequency with DC gate voltage.\label{wcWithgate}}
\end{figure}

A DC voltage on the microwave feedline, allows to tune the coupling capacitance. Such a tunability of the coupling capacitance is best reflected in the shift of the cavity resonance frequency due to its small linewidth. Supplementary Fig.~\ref{wcWithgate} shows the variation in the cavity frequency with DC gate voltage. Based on equivalent lumped circuit parameters and assuming a parallel plate capacitance model, we estimated a displacement of $\sim$~3.1~nm for the graphene drum (as $\Delta x = 2d\frac{C_{sc}}{C_g}\frac{\Delta\omega}{\omega_c}$) at $V_g=+4~V$. The small offset on the gate voltage (-144~mV) likely arises from the work function mismatch between the multilayer graphene and MoRe (gate electrode).

\section{Characterization of graphene resonator:}

\begin{figure}
\begin{center}
\includegraphics[width=140mm]{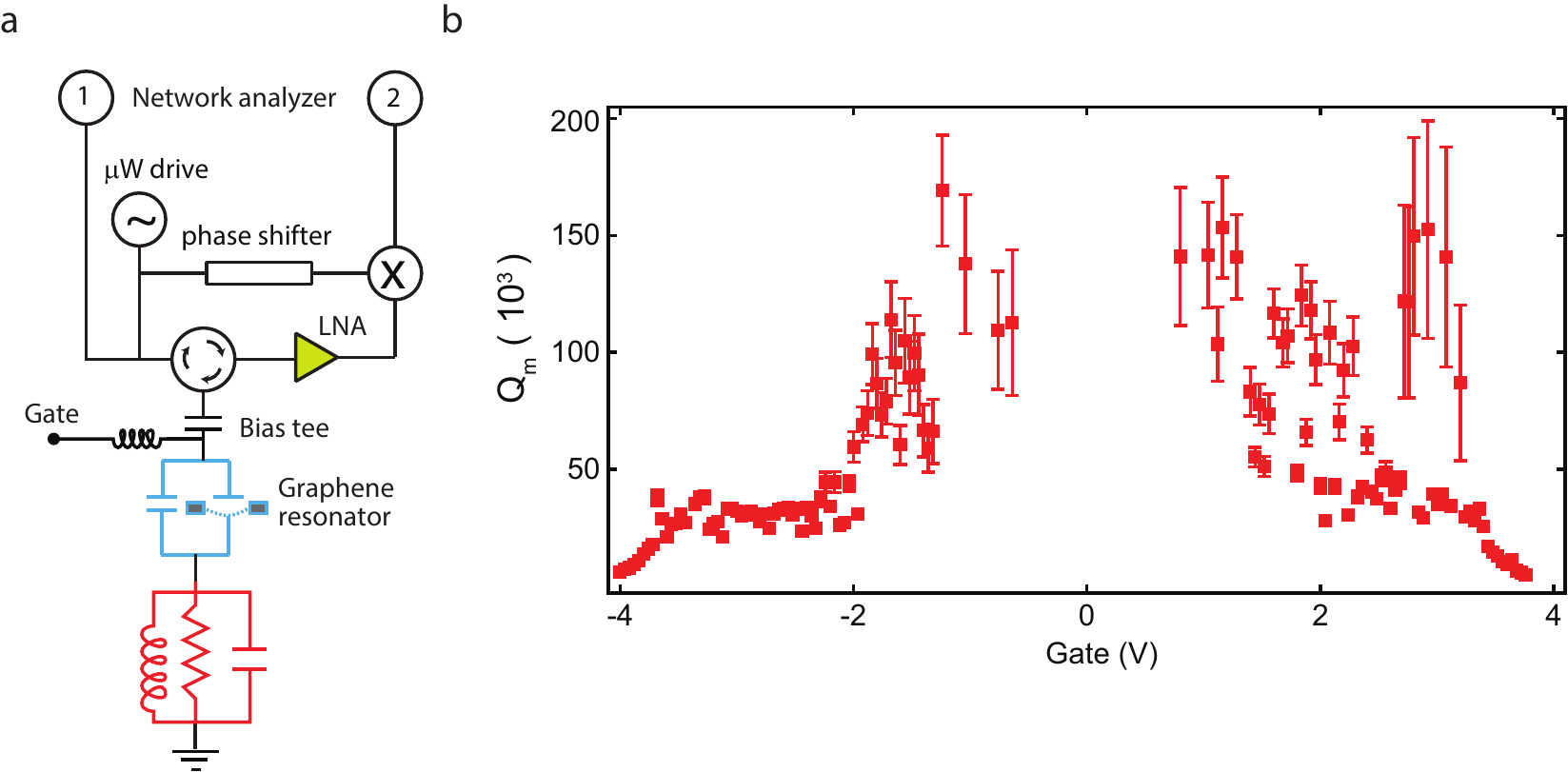}
\end{center}
\caption{(a) Schematic of the homodyne measurement setup to probe the graphene mechanical resonator. (b) Plot of the mechanical quality factor of the graphene resonator. \label{SIfig2}}
\end{figure}

As described in the main text, multilayer graphene dynamics is probed with the homodyne measurement scheme, which allows to measure both quadrature of the mechanical motion. In this scheme, the cavity is driven near its resonance frequency $\omega_c$ (where the slope of any of the quadrature of the cavity response is maximum). The mechanical resonator is driven by an electrostatic drive (using the source of vector network analyzer) by adding a low power RF signal to the feedline and by applying a DC gate voltage. The mechanical driving force is given by $\frac{dC_g}{dx}V_{ac}V_g$. Near the mechanical resonance, the graphene resonator acquires large amplitude. This motion modulates the microwave signal reflected from the cavity and produces signals at frequency $\omega_c\pm\omega_m$. These sidebands along with the carrier signal are routed towards a low noise amplifier using a circulator. At room temperature, we demodulate this signal by mixing it with the phase adjusted carrier reference. The demodulated signal is filtered, amplified and fed into the second port of the vector network analyzer. A measurement of $S_{21}$ in this setup directly probes the responsivity of the mechanical resonator. Using $S_{21}\propto \omega_m^2/(\omega_m^2-\omega^2+i\omega\omega_m/Q_m)$, mechanical resonant frequency $\omega_m$ and quality factor $Q_m$ can be extracted. Supplementary Fig.~\ref{SIfig2}(b) shows the plot of mechanical quality factor with gate voltage, extracted from the data shown in Fig.~2(b) of the main text.

\subsection*{Estimation of the motional amplitude in homodyne measurement scheme :}

Apart from the calibration of the displacement sensitivity (described in the next section), the motional amplitude can also be estimated from the driving force. The ac driving force acting on the resonator ($F_0$, peak) can be related to the peak amplitude by, $x_0 = \frac{F_0Q_m}{m\omega_m^2}$ ($Q_m$ = mechanical quality factor, $m$ = mass, $\omega_m$ = angular resonant frequency) . We have used -100~dBm power (at the gate) to drive the resonator, corresponding to a peak voltage of 3.162~$\mu$V. Using the drum capacitance of 578 aF, 150 nm gap, $V_{dc}$=(150 mV + 144~mV; offset as the frequency dispersion with gate is off-centered by -144~mV), we get a peak amplitude of $\sim$~4~pm.

\section{Calibration of the displacement sensitivity}

For a single port reflection cavity in the resolved sideband limit ($\omega_m \gg\kappa$) with weak sideband detuned drive $\Delta = -\omega_m $ (assuming backaction can be ignored), the displacement sensitivity is given by,
\begin{align*}
S_{xx} &= \frac{1}{2}\left(\frac{\Omega_m}{\eta G}\right)^2 \times \frac{S_N}{P_{io}}
\end{align*}
where, $S_N$ is the microwave noise spectral density and $P_{io}$ is injected microwave drive power at the output of the cavity. Using, the noise temperature of the amplifier $T_{noise}\sim$~2.65~K, a total of 2~dB loss from sample to the HEMT amplifier (circulator, bias tee, non-superconducting coax and two isolators) and taking $P_{io} = -39$~dBm ($n_d\sim7\times10^5$ photons) (detuned to red sideband), we estimated  a displacement sensitivity of $\sim(22$~fm/$\sqrt{\text{Hz}})^2$. The height of the thermal motion peak in the displacement spectral density at 14 mK for this device (in equilibrium with the thermal bath) ideally would be $\sim(17$~fm/$\sqrt{\text{Hz}})^2$, which is smaller than the absolute displacement sensitivity we calculated above.

The displacement sensitivity can be further improved by driving the cavity on resonance ($\Delta=0$). With $\Delta=0$, $S_{xx}$ is given by $\frac{1}{8}\left(\frac{\Omega_m}{\eta G}\right)^2 \times \frac{S_N}{P_{io}}$, which is a factor of 4 improvement (factor of 2 in displacement). However, this increases the intra-cavity photons by an amount $\sim\frac{4\omega_m^2}{k^2}~\sim90000$. Using the on resonance full dynamic range of the cavity (-41~dBm), we get $S_{xx}\sim(17$~fm/$\sqrt{\text{Hz}})^2$. Experimentally we find that this high intra-cavity power leads to the heating of the mechanical mode. Therefore for the calibration of the displacement sensitivity, we chosen to drive the cavity at red sideband and with low power ($n_d\sim7\times10^5$) by which is low enough to avoid any backaction effects and to keep the mechanical mode in equilibrium with the bath.

\begin{figure}
\begin{center}
\includegraphics[width=100mm]{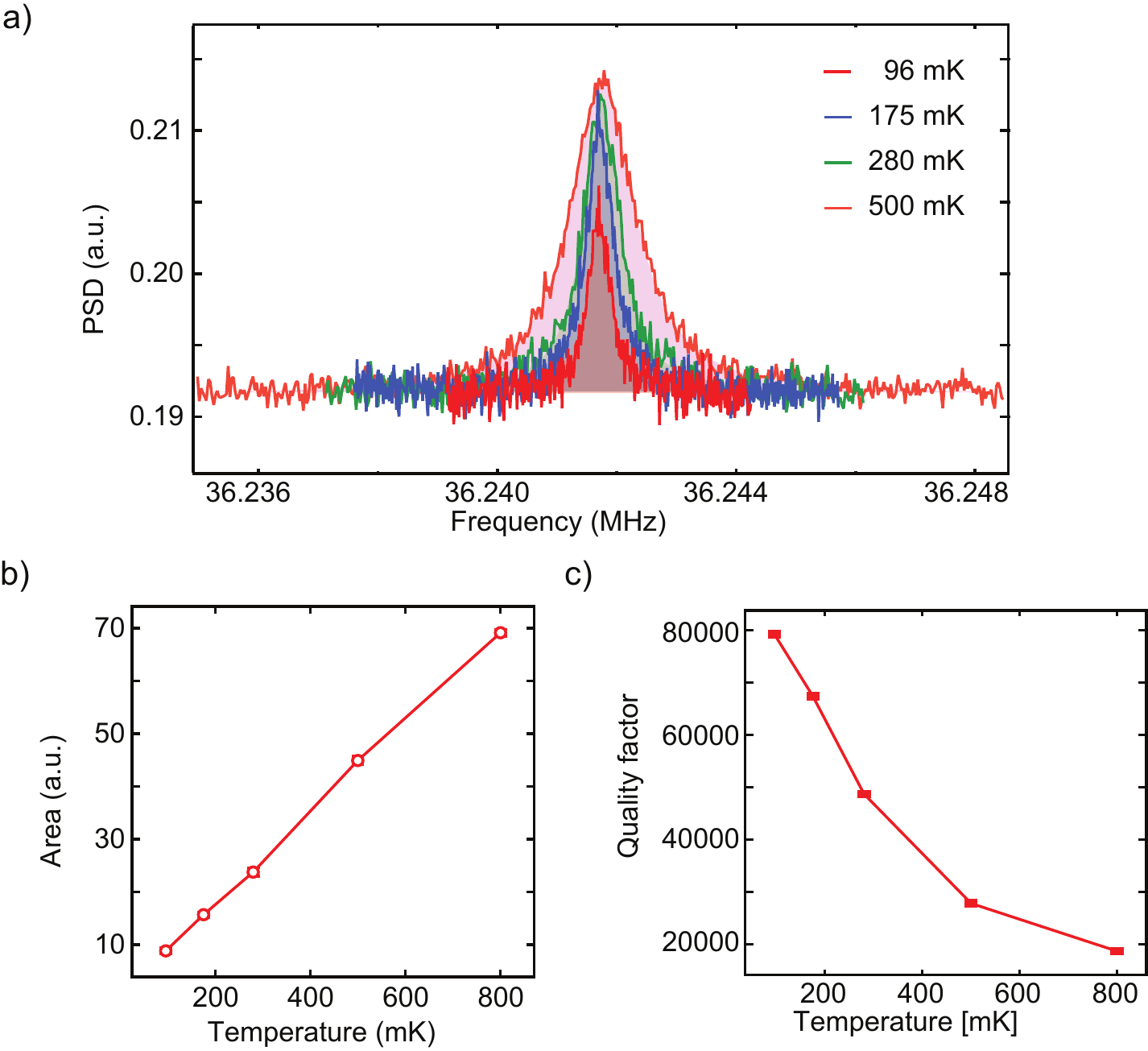}
\end{center}
\caption{a) Power spectral density due to the thermal motion of drum measured by applying a drive at red sideband at different temperatures. The cavity drive frequency has been subtracted from the x axis for clarity. b) The integrated area under the mechanical peak with temperature, showing that mode is thermalized at least down to 96 mK.  c) Change in quality factor with temperature extracted from Lorentzian fits to the thermal measurement data.\label{thermal}}
\end{figure}

\subsection*{Thermal motion of the mechanical resonator :}

As described in the previous section, the better signal to noise ratio (while keeping the mechanical mode still in equilibrium with bath) is achieved by driving the cavity with red detuned microwave. However, sufficiently weak power should be taken to make sure that any back-action effect can be ignored. This limits the overall sensitivity required to resolve thermal motion. By using, $\sim7\times10^5$ intra cavity photons ($P_{io}\sim$-39~dBm, giving $S_{xx}\sim(22$~fm/$\sqrt{\text{Hz}})^2$, we could resolve the thermal motion down to 96 mK. Supplementary Fig.~\ref{thermal} (a) shows the measurement of the microwave power spectral density (PSD) measured at the cavity frequency while driving the cavity at red sideband ($\gamma_{dba} < 0.04\times \gamma_m$) at different temperatures. Measurement setup involves driving the cavity on the red sideband and directly measuring the up-converted sideband appearing at $\omega_c$. The microwave PSD is measured by typically taking 1001 points in the measurement span and performing upto $\sim$10000 power trace averages in FFT mode of the spectrum analyzer (RBW=5 Hz, VBW=50 Hz). The cavity drive frequency from the x-axis in Supplementary Fig.~\ref{thermal} (a) has been subtracted for clarity.

To demonstrate the thermalization of the mechanical mode with the phonon-bath, we have plotted the integrated area under the thermal peak for different temperatures in Supplementary Fig.~\ref{thermal}. A linear dependence shows the mode thermalization down to 96~mK. Supplementary Fig.~\ref{thermal}(c) shows the plot of the mechanical quality factor extracted by fitting a Lorentzian to the microwave PSD measurements.

\subsection*{Displacement sensitivity and standard quantum limit :}

For a continuous linear detector such as here (a single port reflection cavity), we compare the displacement sensitivity with that of due to the standard quantum limit (SQL). The displacement spectral density at SQL is given by $S_{xx}^{SQL}=\frac{2x_{zpf}^2}{\gamma_m}$. Assuming the total measurement chain adds $n_{added}$ quanta of noise, in sideband resolved limit, for on resonance drive of the cavity ($\Delta = 0$), we get
\begin{align}
\frac{S_{xx}^{imp}}{S_{xx}^{SQL}} = \frac{n_{added}}{\eta}\frac{1}{C} \left(\frac{\omega_m}{\kappa}\right)^2.
\end{align}
where, $C$ is defined as $\frac{4g_0^2 n_d}{\gamma_m\kappa}$ and $n_d$ is the number of photons inside the cavity. Using the full dynamic range of the cavity (-41~dBm), which also reduces the mechanical quality factor to 18000, we estimated $\frac{S_{xx}^{imp}}{S_{xx}^{SQL}}\sim~(107)^2$, which is way above the SQL. It is evident from above equation that the main reason for such a high ratio is the extreme sideband resolution of our device $\frac{\omega_m}{\kappa}\sim~150$.

It is interesting to work out this ratio for sideband detuned drives $(|\Delta| = \omega_m)$. The effect of backaction on the mechanics due to sideband detuning, can be very well captured by replacing the mechanical dissipation rate by $(1\pm C)\gamma_m$ (where, positive (negative) sign is taken for red (blue) detuned drive.). For a sideband detuned drive, we get
\begin{align}
\frac{S_{xx}^{imp}}{S_{xx}^{SQL}} = \frac{n_{added}}{\eta}\frac{1\pm C}{C}.
\end{align}
Therefore, for a red detuned drive, it is only possible to reach the SQL with a quantum limited amplifier and extremely overcoupled cavity ($\eta = 1$) which will add only half a quanta of noise (cooperativity ratio drops out with large $C$). With a blue detuned drive, however, one can asymptotically approach $C$ to 1 (before the self oscillation of the mechanical resonator), reducing the $(1-C)$ to close to zero and making a measurement with a precision well below set by the standard quantum limit. 
This comes about from an optomechanical feedback which reduces the effective dissipation rate of the mechanical mode, increasing the spectral density of zeropoint fluctuations. This is also accompanied by optomechanical gain, which will amplify the motion of the mode, along with any accompanying thermal noise it might posses.

\section{Calibration of single photon coupling strength $g_0$ using frequency modulation technique:}

The single photon coupling strength can be calibrated by  sending a frequency modulated (FM) signal to the cavity, provided that the mechanical resonator is in thermal equilibrium with the bath \cite{gorodetsky_determination_2010,zhou_slowing_2013}. By driving the cavity with an FM signal with a known modulation index $\phi_0$ at modulation frequency $\Omega_{mod}$ (near $\omega_m$ but outside mechanical linewidth), the single photon coupling strength can be extracted by 
\begin{align}
g_0^2 = \frac{1}{2\bar{n}_{th}}\frac{\phi_0^2 \omega_m^2}{2}\frac{S_{VV}(\omega_m)}{S_{VV}(\pm\Omega_{mod})}*\frac{\gamma_m/4}{ENBW}
\end{align}
where, $S_{VV}(\omega_m)$ is power spectral density measured at the mechanical frequency, $S_{VV}(\Omega_{mod})$ is PSD measured at modulation frequency, and mechanical dissipation rate $\gamma_m$ and effective noise bandwidth (ENBW) are taken in direct frequency units. Effectively, the product of last two fractions is the ratio of area under the mechanical peak to the area under the modulation peak. 

\begin{figure}
\begin{center}
\includegraphics[width=100mm]{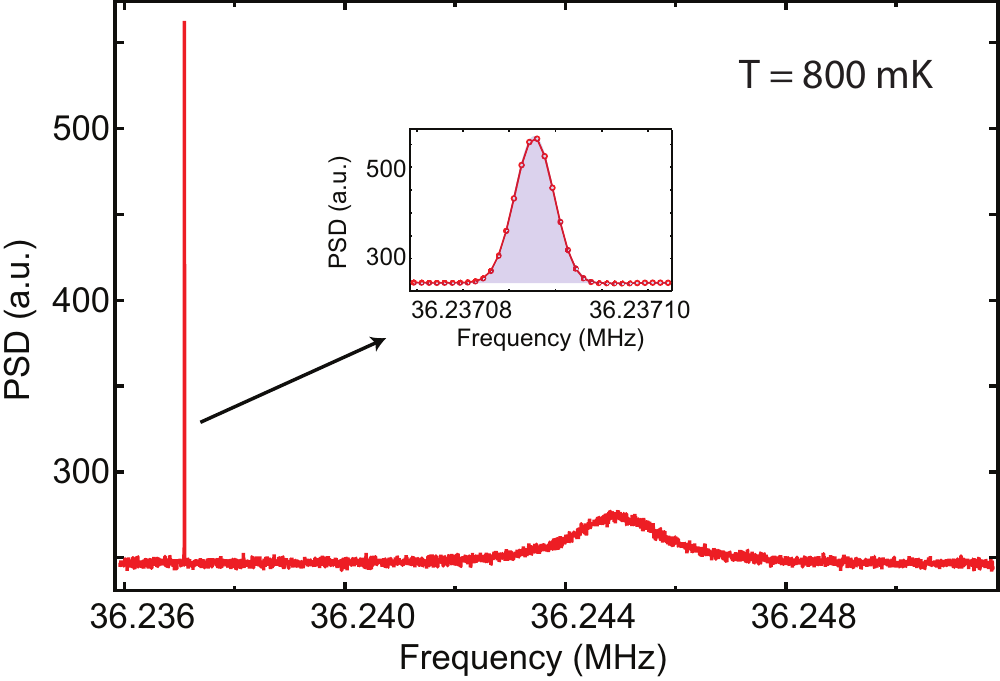}
\end{center}
\caption{ Power spectral density due to the thermal motion of drum measured by applying a frequency modulated signal with a carrier frequency detuned to red sideband ($T$ = 800 mK).\label{FMCalib}}
\end{figure}

To generate a frequency modulation signal near mechanical frequency ($\sim$~36 MHz), we use Rohde $\&$ Schwarz SGS100A vector source. We drive the cavity with low power signal at red sideband (so backaction effects can be neglected) with a modulation depth $\phi_0 = 8.6\times10^{-7}$ and a modulation frequency near mechanical resonance and measure the power spectral density collected near the cavity frequency. The modulation depth was calibrated by measuring the voltage ratio of the first sideband signal to the carrier signal, which follows $J_1(\phi_0)/J_0(\phi_0)$, which for $\phi_0\ll 1$ can be approximated as $\phi_0/2$. Supplementary Fig.~\ref{FMCalib} shows such a measurement of noise spectral density measured at 800 mK. The inset shows the detailed view of modulation peak. By fitting to modulation peak to a Gaussian (filter convolution function) and mechanical peak to a Lorentzian, we calculated the area under the curves, which gives us single photon coupling strength of $g_0 = 2\pi\times0.83$~Hz. Together with the OMIA and OMIR measurements, calibration of $g_0$ allows us to calibrate the number of photons inside the cavity.
 
\section{Optomechanical interaction :}

\subsection*{Optomechanical induced absorption and reflection}

To explain the optomechanical effects on the cavity, we have followed the approach taken in the supporting online material of ref~ \cite{weis_optomechanically_2010}. Our starting point is the expression for the field amplitude $A^{-}$ inside the cavity at the probe frequency given by
\begin{align}
A^-=-\sqrt{\kappa_e}S_p\frac{1+if(\Omega)}{i(\bar{\Delta}+\Omega)-\kappa/2-2\bar{\Delta}f(\Omega)},
\end{align}
where
\begin{align}
f(\Omega)=\hbar G^2\bar{a}^2\frac{\chi(\Omega)}{i(\bar{\Delta}-\Omega)+\kappa/2} 
\end{align}
and
\begin{align}
\chi(\Omega)=\frac{1}{m_{eff}}\frac{1}{\omega_m^2-\Omega^2-i\Omega\gamma_m}.
\end{align}
Here $\kappa_e$ is the external decay rate, $\kappa$ is total cavity decay rate, $S_p$ is the probe amplitude, $G$ is the cavity pull-in factor, $\bar{a}$ is the drive field amplitude, $\Omega=\omega_p-\omega_d$ is the beating frequency, $\bar{\Delta}=\omega_d-\omega_c$ is the detuning of the drive signal from the cavity center frequency, $\omega_m$ is the resonance frequency of the mechanical resonator, $m_{eff}$ is the effective mass of the resonator, and $\gamma_m$ is mechanical dissipation rate.

To relate $A^-$ to the reflection coefficient of the cavity, we use the input-output relation for a single port cavity, defining
\begin{align}
S_{11} = 1 - \sqrt{\kappa_e}\frac{A^-}{S_p} .
\end{align}

For a \textbf{red} detuned drive, we take $\bar{\Delta}=-\omega_m$ and $\Omega\simeq
\omega_m$. We introduce $\Delta^{'}=\Omega-\omega_m$ as a small detuning parameter. Taking a resolved sideband approximation $\kappa\ll\omega_m$, we get
\begin{align}
\chi(\Omega)= - \frac{1}{2m_{eff}\omega_m}\frac{1}{\Delta^{'}+i\gamma_m/2}
\end{align}
and
\begin{align}
f(\Omega)=-\frac{ig^2}{8\omega_m}\frac{1}{\Delta^{'}+i\gamma_m/2}
\end{align}
where we have used $g=G\bar{a}\sqrt{\frac{\hbar}{2m_{eff}\omega_m}}$.

For a \textbf{blue} detuned drive, we take $\bar{\Delta}=\omega_m$ and $\Omega\simeq\omega_m$, introducing $\Delta^{'}=\Omega+\omega_m$ as a small detuning parameter. This gives
\begin{align}
\chi(\Omega)= - \frac{1}{2m_{eff}\omega_m}\frac{1}{\Delta^{'}+i\gamma_m/2}
\end{align}
and
\begin{align}
f(\Omega)=-\frac{ig^2}{8\omega_m}\frac{1}{\Delta^{'}+i\gamma_m/2}.
\end{align}

By introducing normalized variables,
$\eta=\frac{\kappa_e}{\kappa}$, $g^{'}=\frac{g}{\kappa}$,$\gamma=\frac{\gamma_m}{\kappa}$, $\omega=\frac{\omega_p-(\omega_d\pm\omega_m)}{\kappa}$ and $\delta=\frac{\omega_c-(\omega_d\pm\omega_m)}{\kappa}$  
the reflection coefficient of the cavity can be written in the following form
\begin{align}
S_{11}(\omega,\delta)=1-\frac{i\eta(-2i\gamma-4\omega)}{\pm4g^{'2}+(-i+2\delta-2\omega)(i\gamma+2\omega)}\label{eq:norm}
\end{align}
where positive (negative) sign is taken for red (blue) detuned drives.

\subsection*{Backaction on the cavity within the mechanical bandwidth}

In the limit of exact detuning, following equation~(\ref{eq:norm}), the magnitude of the cavity reflection at $\omega_p=\omega_c$ can be written as
\begin{align}
|S_{11}(\omega=0,\delta=0)|=\left| 1- 2\frac{\kappa_e}{\kappa_e+(\kappa_i\pm\frac{4g^2}{\gamma_m})}\right|.\label{eq:limit}
\end{align}
Comparing it with the reflection coefficient of the cavity in absence of the optomechanical coupling \textit{i.e.} $|S_{11}(\omega_p=\omega_c)|=\left|1-2\frac{\kappa_e}{\kappa_e+\kappa_i}\right|$, it becomes quite evident that within the mechanical bandwidth the effect of the optomechanical coupling can be captured by modifying internal dissipation rate to $\kappa_i\pm\frac{4g^2}{\gamma_m}$.

\subsection*{Comparison between single port reflection cavity and cavity side coupled to a transmission line:}

\begin{figure}
\begin{center}
\includegraphics[width=140mm]{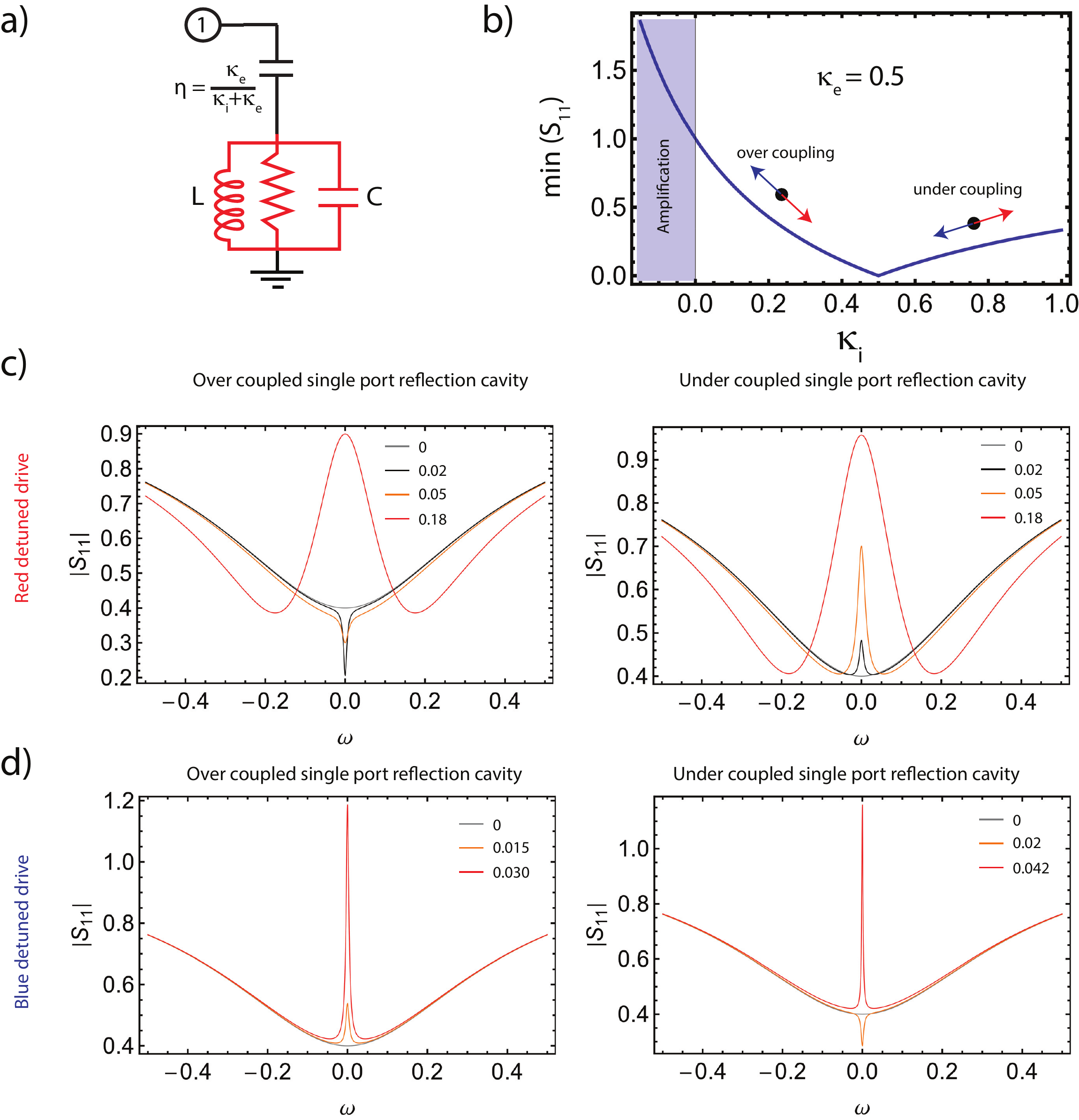}
\end{center}
\caption{(a) Schematic representation of a single port reflection cavity. (b) Plot of minimum of $S_{11}$ for a single port reflection cavity with varied internal dissipation rate $\kappa_i$ and fixed external dissipation rate $\kappa_e$. The effect of optomechanical coupling can be captured by modifying the internal dissipation rate. Red detuned drive takes the cavity towards undercoupling eventually leading to normal mode splitting. A blue detuned drive modify it towards the overcoupling eventually taking the system to amplification via the mechanical resonator. (c, d) Calculated plots of the reflection coefficient $S_{11}$ for single port reflection cavity with varied coupling strength. (c) With red detuned drive for an over-coupled $\eta > 0.5$ (left) and under-coupled cavity $\eta < 0.5$ (right).  (d) With a blue detuned drive for an over-coupled $\eta > 0.5$ (left) and under-coupled cavity $\eta < 0.5$ (right). \label{Single}}
\end{figure}

\begin{figure}
\begin{center}
\includegraphics[width=140mm]{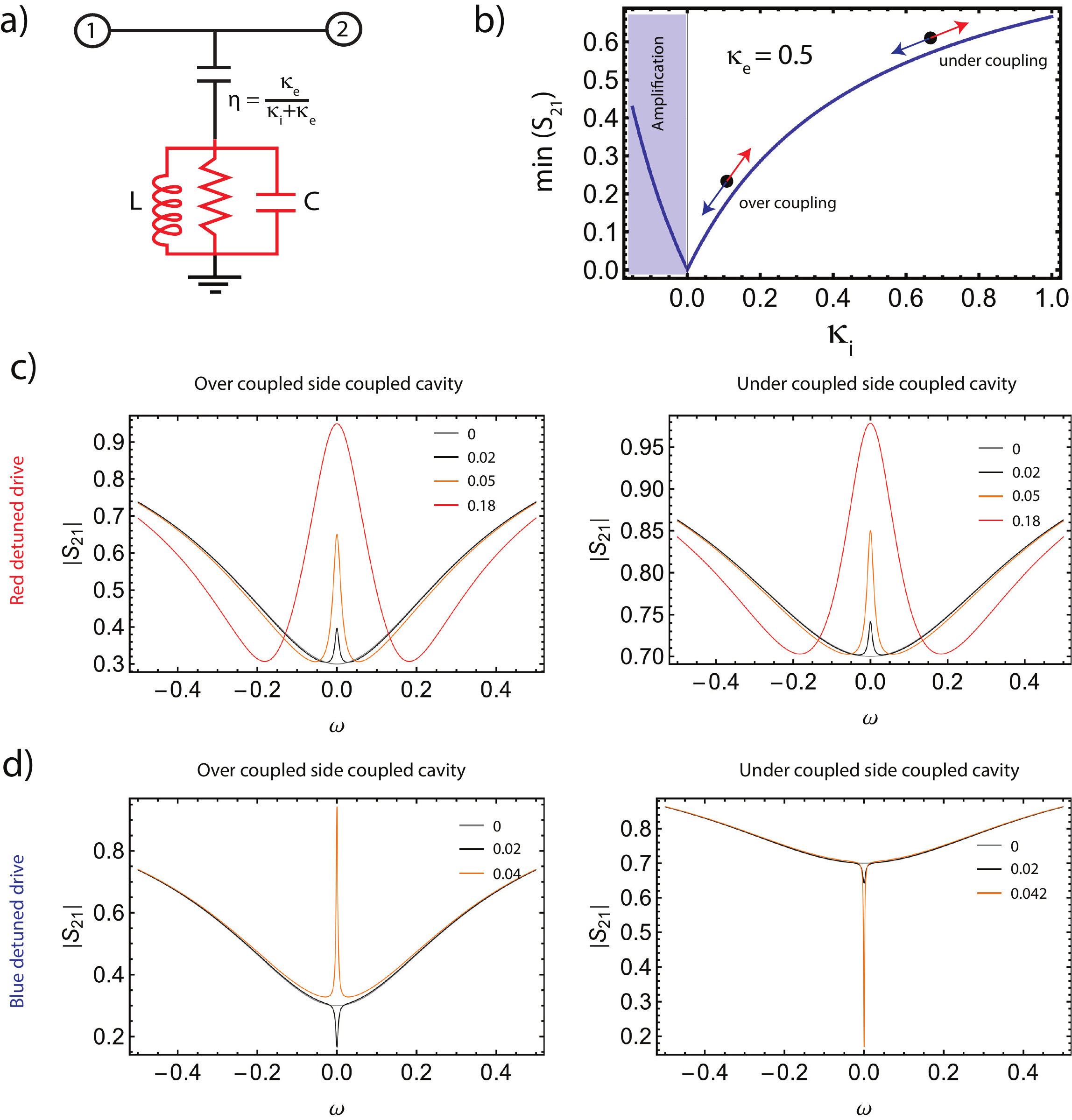}
\end{center}
\caption{(a) Schematic representation of a side coupled transmission cavity. (b) Plot of minimum of $S_{21}$ for a side coupled cavity with varied internal dissipation rate $\kappa_i$ and fixed external dissipation rate $\kappa_e$. The effect of optomechanical coupling can be captured by modifying the internal dissipation rate. Red detuned drive takes the cavity towards undercoupling eventually leading to normal mode splitting. A blue detuned drive modify it towards the overcoupling eventually taking the system to amplification via the mechanical resonator. (c, d) Calculated plots of the transmission coefficient $S_{21}$ for side coupled cavity with varied coupling strength. (c) With red detuned drive for an over-coupled $\eta > 0.5$ (left) and under-coupled cavity $\eta < 0.5$ (right).  (d) With a blue detuned drive for an over-coupled $\eta > 0.5$ (left) and under-coupled cavity $\eta < 0.5$ (right).\label{Side}}
\end{figure}

In this section, we compare differences between a single port reflection cavity (studied here) to  a side coupled cavity coupled to a two port transmission line. The representative examples are shown in Supplementary Fig.~\ref{Single}(a) and Supplementary Fig.~\ref{Side}(a). In the previous section, we calculated the reflection coefficient for a single port reflection cavity with optomechanical coupling. In the absence of any optomechanical interaction, the minimum of $S_{11}$ as a function of coupling fraction $\eta$, for a fixed $\kappa_e = 0.5$, is plotted in Supplementary Fig.~\ref{Single}(b). At critical coupling $\eta=0.5$, $S_{11}$ becomes zero and with further increase in $\eta~( > 1)$, we get amplification (as $S_{11} > 1$). The optomechanical interaction as pointed out earlier, can be captured with a modified the internal dissipation rate $\kappa_i\rightarrow\kappa_i\pm\frac{4g^2}{\gamma_m}$, where positive (negative) sign is taken for red (blue) detuned drive. Therefore effectively, within the mechanical linewidth, red detuned drive takes the $S_{11}$ towards an undercoupled limit (by making the effective internal dissipation rate larger). Hence, for an over coupled cavity (to begin with), at low optomechanical coupling an absorption feature (optomechanical induced absorption OMIA) appears in the cavity response. Once optomechanical coupling becomes sufficiently larger, this OMIA feature converts into optomechanical induced reflection peak. This is also called the "onset of the normal mode splitting" as illustrated in the left panel of Supplementary Fig.~\ref{Single}(c). However, for an undercoupled cavity (to begin with), optomechanical interaction with red detuned drive can only make it more and more undercoupled,eventually leading to the normal mode splitting. This has been illustrated in the right panel of Supplementary Fig.~\ref{Single}(c).

The blue detuned drive, however, makes the response appear more like that of an overcoupled system (as effective internal dissipation rate gets reduced by $\kappa_i\rightarrow\kappa_i-\frac{4g^2}{\gamma_m}$ ). Therefore for an overcoupled cavity to begin with, optomechanical interaction lead to an optomechanical $S_{11} > 1$ as illustrated in the left panel of Supplementary Fig.~\ref{Single}(d). However for an undercoupled cavity to start with, at low optomechanical coupling strengths, an OMIA features appears, which at sufficiently large coupling strength converts into OMIR feature and eventually into amplification of the probe photons as shown in the right panel of the Supplementary Fig.~\ref{Single}(d).

Now, we contrast this behavior with the cavity side coupled to a transmission line as shown in the Supplementary Fig.~\ref{Side}(a). Since only half of the intracavity field couples back to the transmitted field, the transmission coefficient $S_{21}$ can be written as, 
\begin{align}
S_{21}(\omega,\delta)=1-\frac{1}{2}\times\frac{i\eta(-2i\gamma-4\omega)}{\pm4g^{'2}+(-i+2\delta-2\omega)(i\gamma+2\omega)}\label{eq:norm}
\end{align}
where positive (negative) sign is taken for red (blue) detuned drives.

Supplementary Fig.~\ref{Side}(b) shows the plot of the minimum of $S_{21}$ with varied coupling strength to the transmission line. In contrast to single port reflection cavity, we see only a monotonous behavior of $min(S_{21})$ for $0<\eta<1$. As mentioned earlier, the effect of optomechanical coupling can be captured by taking an effective internal dissipation rate $\kappa_i\rightarrow\kappa_i-\frac{4g^2}{\gamma_m}$. Therefore, for a red detuned drive, within the mechanical line width, optomechanical interaction makes the cavity response more and more under coupled which appears as an increased in $min(S_{21})$, and hence an OMIR like feature. For higher optomechanical coupling strength, this OMIR feature evolves into the normal mode splitting. This has been shown for different optomechanical coupling strengths in Supplementary Fig.~\ref{Side}(c) for overcoupled (left) and undercoupled (right) cases. 

However for a blue detuned drive, optomechanical interaction reduces the effective internal dissipation rate of the cavity and within mechanical linewidth reduces the $S_{21}$ response, appearing as an OMIA feature. For sufficiently large optomechanical coupling strength, effective internal dissipation rate $\kappa_i-\frac{4g^2}{\gamma_m}$, can become negative leading to the amplification of the probe photons. This has been illustrated in Supplementary Fig.~\ref{Side}(d) for an overcoupled and undercoupled coupling strength to the transmission line.

To summarize, the response of a single port reflection cavity can show OMIA feature with red detuned drive, if it is overcoupled to start with. Whereas, a side coupled cavity, irrespective to its coupling strength to the transmission line, will always show an OMIR feature with red detuned drive.

\subsection*{Measurement of the mechanical quality factor and estimation of the group delay at $V_g~=~0$~V}

\begin{figure}
\begin{center}
\includegraphics[width=140mm]{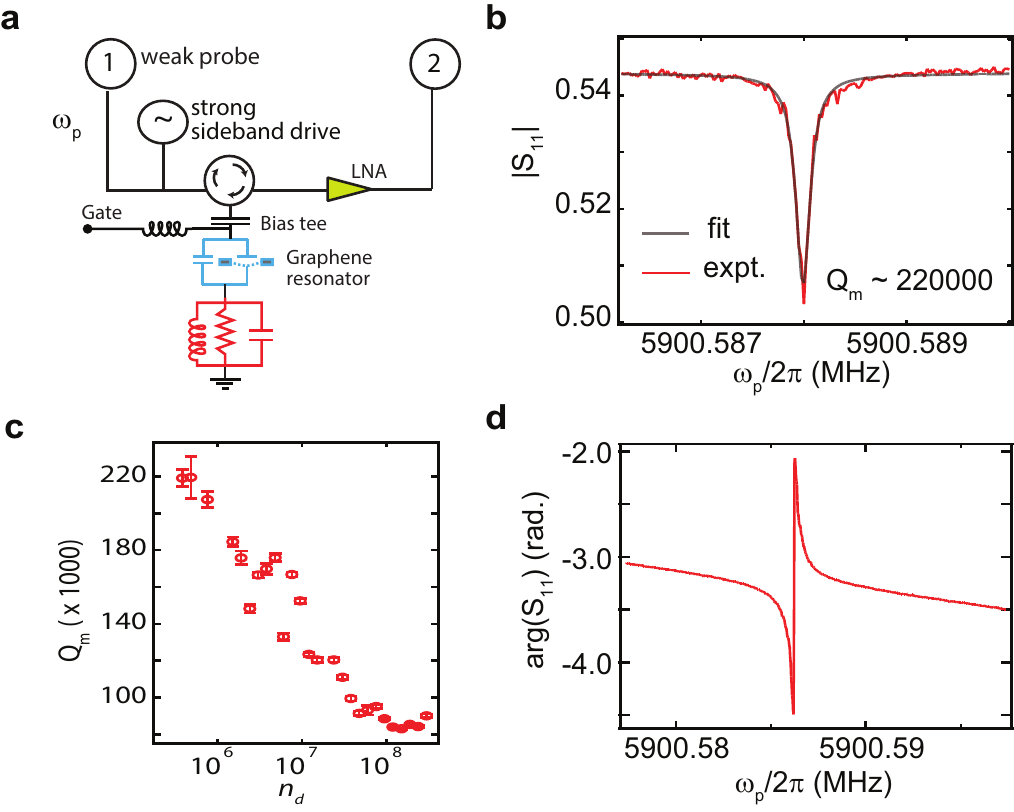}
\end{center}
\caption{(a) Measurement scheme to probe optomechanical interaction. (b) Detailed measurement of OMIA feature measured with low optomechanical coupling strength (red) along with the fitted curve (gray). (c) Plot of the mechanical quality factor with red detuned intracavity photon number $n_d$ extracted from the OMIA/OMIR data. (d) Plot of phase of the reflection coefficient measured with red detuned drive. The phase swing in the mechanical bandwidth corresponds to a group delay of $\approx$10~ms.\label{SIfig3}}
\end{figure}

Supplementary Fig.~\ref{SIfig3}(a) shows the schematic of the measurement setup used to study the optomechanical interactions. A weak probe signal is applied using a vector network analyzer. A separate microwave source is used to drive the cavity near the sidebands. The two signal tones are added using a directional coupler (10~dB). Supplementary Fig.~\ref{SIfig3}(b) shows the measurement of $|S_{11}|$ with small optomechanical coupling strength such that any backaction effect on the mechanical resonator can be ignored (red curve). This can be fit to equation~(\ref{eq:norm}) to yield a mechanical quality factor $Q_m\approx$~220,000 at $V_g$=0~V. Supplementary Fig.~\ref{SIfig3}(c) shows the mechanical quality factor extracted from the OMIA/OMIR measurements at different number of drive photons. 
At our largest cooperativity $C=8$, the quality factor of $\sim$~88000. From the temperature dependence of the quality factor, we estimate a mode temperature of $\sim$~100~mK. The reason for the increased mode temperature is not certain, but it could arise either from heating of the mechanical mode by the non-thermalized photon field or from resistive losses in the graphene sheet. In future experiment, these possibilities can be investigated with sufficient microwave attenuations and with the use of superconducting two dimensional exfoliated crystals. 
This reflects in the sub-linear dependence of cooperativity with $n_d$ as shown in the Figure~4~(d) of the main text.   

Supplementary Fig.~\ref{SIfig3}(d) shows the phase response $\phi$ of the cavity at somewhat larger coupling strength such that $\kappa_e\simeq\kappa_i+\frac{4g^2}{\gamma_m}$. The swing in the phase can be associated with a group delay $\tau=\frac{\delta\phi}{\delta\Omega}$ of the photons which lie within the mechanical bandwidth at $\omega_c$. From this measurement, we estimated a maximum group delay of $\approx$~10~ms. It should be noted that, such a group delay is also accompanied by large loss of the signal, which has been shown in Supplementary Fig.~\ref{SIfig4}.

\section{Estimation of the cooperativity :}

\begin{figure}
\begin{center}
\includegraphics[width=140mm]{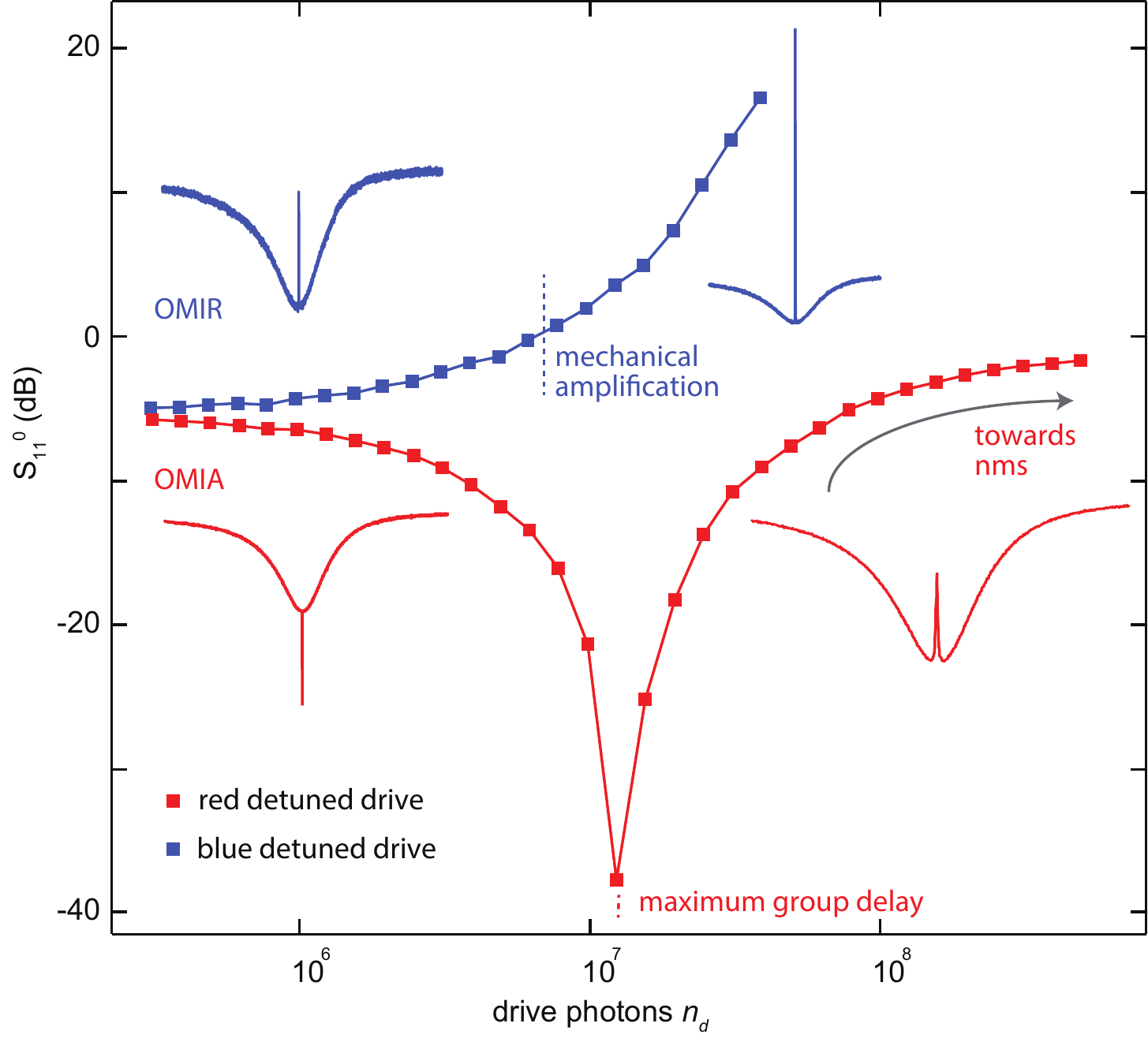}
\end{center}
\caption{Plot of the reflection coefficient measured at cavity resonance frequency with zero detuning $S_{11}^0\equiv S_{11}(\omega=0,\delta=0)$ of the red and blue detuned drive. Regions with different effects have been marked.\label{SIfig4}}
\end{figure}
Using the notion of the optomechanical cooperativity $C=\frac{4g^2}{\kappa\gamma_m}$, equation~(\ref{eq:limit}) can be written as,
\begin{align}
S_{11}^0\equiv|S_{11}(\omega=0,\delta=0)|=\left| 1- \frac{2\eta}{1\pm C}\right|\label{eq:C}
\end{align}
allowing a direct estimation of cooperativity without any free parameters, where positive (negative) sign is taken for red (blue) detuned drive.
 
Supplementary Fig.~\ref{SIfig4} shows the plot of $|S_{11}^0|$ with the number of drive photons for the red and blue detuned drives. For a blue detuned drive while increasing the coupling strength (increase in the number of drive photons), the optomechanical induced reflection (OMIR) feature converts into mechanical amplification ($S_{11}^0 (dB) > 0$). This cross over occurs when internal losses in the cavity are balanced by $4g^2/\gamma_m$. For a red detuned drive with increase in the coupling, the optomechanical induced absorption first becomes deeper and then starts to convert into OMIR like feature. This cross over occurs when external losses are balanced by $\kappa_i+4g^2/\gamma_m$. Any further increase in coupling leads the system towards normal mode splitting (NMS) of the cavity resonance (strong coupling limit). Following this measurement, we can estimate the cooperativity $C$ by using equation~(\ref{eq:C}). The result has been shown in the Figure~4(d) of the main text. 

\subsection*{Scaling of cooperativity with the area of the resonator}

The relevant figure of merit for reaching the quantum coherent regime is when cooperativity becomes equal to the number of thermal phonon in equilibrium with the phonon bath. This ratio can be written as,
\begin{align*}
\frac{C}{\bar{n}_{th}} = \frac{4G^2n_dx_{zpf}^2}{\gamma_m\kappa} \frac{\hbar\omega_m}{k_BT} = \left(\frac{2\hbar^2n_d}{k_BT \kappa}\right)\left(\frac{G^2Q_m}{m\omega_m}\right)
\end{align*}
By making larger mechanical resonators, in plate limit the resonance frequency is expected to scale as $\frac{t}{A}$, where $t$ is the thickness of the plate and $A$ is the area of the drum. 
The mass of the resonator also scale linear with $A$.
As long as the graphene mechanical capacitor does not dominate the total capacitance of the cavity, the cavity pull-in 
parameter $G$ also scales linearly with area. 
For the scaling, we will assume that the mechanical quality factor stay constant upon making larger area drums. 
Doing so, we come to a scaling of $C/\bar{n}_{th}$ that goes as 
$(A/t)^2$:
\begin{center}
$\frac{C}{\bar{n}_{th}} \propto \frac{A^2}{t^2} \mathrm{(plate\ limit)} $
\end{center}

For the case that tension dominates over the bending rigidity, often seen for single-layer graphene, the mechanical resonance frequency will scale as $\frac{1}{\sqrt{At}}$, implying that for fixed mechanical quality factor $\frac{C}{\bar{n}_{th}}$ will scale as $\sqrt{A^3/t}$:
\begin{center}
$\frac{C}{\bar{n}_{th}} \propto \frac{A^{3/2}}{t^{1/2}} \mathrm{(tension\ limit)}$
\end{center}

We also note that the quality factor of graphene resonators often increases for larger area drums, suggesting that the scaling of $C/\bar{n}_{th}$ with area could even be more favorable than the estimates above.